\documentclass[letterpaper]{article}
\pdfoutput=1
\usepackage{jheppub}
\usepackage{graphicx, caption, subcaption}  
\usepackage{dcolumn}   
\usepackage{amsmath, amssymb}        
\usepackage{color,latexsym, array,multirow,  verbatim, enumerate,cancel}

\def\issue(#1,#2,#3){{\bf #1}, #2 (#3)}

\catcode`\@=11
\def\lsim{\mathrel{\mathpalette\@versim<}}
\def\gsim{\mathrel{\mathpalette\@versim>}}
\def\@versim#1#2{\vcenter{\offinterlineskip
\ialign{$\m@th#1\hfil##\hfil$\crcr#2\crcr\sim\crcr } }}
\catcode`\@=12

\catcode`@=12

\newcommand{\met}{$\cancel E_T$}

\newcommand{\newc}{\newcommand}
\newc{\wt}{\widetilde}
\newc{\ra}{\rightarrow}

\def\beq {\begin{equation}}
\def\eeq {\end{equation}}
\def\bi {\begin{itemize}}
\def\ei {\end{itemize}}
\def\bea {\begin{eqnarray}}
\def\eea {\end{eqnarray}}
\def \PMET{\rm p{\!\!\!/}_T}

\def \met{\rm E{\!\!\!/}_T}

\newcommand{\br}{\begin{eqnarray}}
\newcommand{\er}{\end{eqnarray}}
\newcommand{\be}{\begin{equation}}
\newcommand{\ee}{\end{equation}}

\newcommand{\ch}{\widetilde \chi^\pm}

\def\lum             {{\cal L}}
\newcommand{\ifb} {\rm {fb}^{-1}}

\def \chii {{\wt\chi_i}}
\def \chij {{\wt\chi_j}}

\def \chonep {{\wt\chi_1^+}}

\def \ch2p {{\wt\chi_2^+}}
\def \chonem {{\wt\chi_1^-}}
\def \ch2m {{\wt\chi_2^-}}

\def \chip{{\wt\chi_i}^{+}}

\def \chjm{{\wt\chi_j}^{-}}

\def \chonepm{{\wt\chi_1}^{\pm}}
\def \chonemp{{\wt\chi_1}^{\mp}}

\newc{\dmchi}{\Delta m_{\wt\chi}}
\def \chtwop {{\wt\chi_2^+}}
\def \chtwom {{\wt\chi_2^-}}
\def \chtwopm{{\wt\chi_2}^{\pm}}
\def \chtwomp{{\wt\chi_2}^{\mp}}

\def \chkpm{{\wt\chi_k}^{\pm}}

\def \chlmp{{\wt\chi_\ell}^{\mp}}

\def \lspi{\wt\chi_i^0}

\def \lspj{\wt\chi_j^0}

\def \lspone{\wt\chi_1^0}
\def \mlspone{m_{\lspone}}
\def \lsptwo{\wt\chi_2^0}

\def \lspthree{\wt\chi_3^0}

\def \lspfour{\wt\chi_4^0}

\def\issue(#1,#2,#3){{\bf #1}, #2 (#3)}

\hypersetup{
   colorlinks=true,       
   linkcolor=blue,        
   citecolor=red,         
   filecolor=magenta,      
   }

\title{Study of MSSM heavy Higgs bosons decaying into charginos and neutralinos }

\author[a]{Rahool K. Barman,}
\author[a]{Biplob Bhattacherjee,}
\author[b]{Amit Chakraborty,}
\author[c,d]{and Arghya Choudhury}
\affiliation[a]{Centre for High Energy Physics, \\ 
Indian Institute of Science, Bangalore 560012, India}
\affiliation[b]{Department of Theoretical Physics,  
Tata Institute of Fundamental Research,\\
1, Homi Bhabha Road, Mumbai 400005, India}
\affiliation[c]{Consortium for Fundamental Physics, Department of Physics and Astronomy, \\
University of Sheffield, Sheffield S3 7RH, United Kingdom}
\affiliation[d]{Consortium for Fundamental Physics, Department of Physics and Astronomy, \\
University of Manchester, Manchester, M13 9PL, United Kingdom}
\emailAdd{rahoolkbarman@chep.iisc.ernet.in}
\emailAdd{biplob@chep.iisc.ernet.in}
\emailAdd{amit@theory.tifr.res.in}
\emailAdd{a.choudhury@sheffield.ac.uk}
\date{\today}
\abstract
{ A multitude of searches have already been performed by the ATLAS and CMS 
collaborations at the LHC to probe the heavy Higgses of the Minimal 
Supersymmetric Standard Model (MSSM) through their decay to the 
Standard Model particles. In this paper, we study the decay of the 
MSSM heavy Higgses into
neutralino and chargino pairs and estimate the maximum possible
branching ratios for these `ino' modes being consistent with the
present LHC data. After performing a random scan of the relevant 
electroweakino parameters,
we impose the SM 125 GeV Higgs constraints, low energy flavour data as well
as current bounds on heavy Higgses from the LHC run-I and run-II data.
The present limits on the electroweakino masses and couplings are also
considered in our analysis. We choose a few representative benchmark
points satisfying all the above-mentioned constraints, and then
perform a detailed collider simulation, including fast detector effects,
and analyze all the potential SM backgrounds in order to estimate the discovery
reach of these heavy Higgses at the LHC. We restrict ourselves within
the leptonic cascade decay modes of these heavy Higgses and study the
mono-X + $\met$ (X=W,Z) and trilepton + $\met$ signatures in the context of high
luminosity run of the 14 TeV LHC. 
}

\begin{document}
\maketitle


\newpage
\section{Introduction}
\label{intro}

The observation of a resonance around 125 GeV by the ATLAS and CMS collaborations 
has led to a new era in particle physics \cite{atlas_125,cms_125}. 
Comprehensive studies to investigate the spin and parity quantum 
numbers of the observed particle prefer its scalar nature. Various 
properties of this newly discovered resonance also seem to be in good accordance 
with that of the Standard Model (SM) Higgs boson. However, it is to be noted 
that even though large deviations from the SM predictions have already 
been excluded from the current LHC data, non-standard Higgs couplings 
are still allowed within the present uncertainties in various Higgs 
coupling measurements \cite{ATLAS-CMS-comb,Aad:2015gba,Khachatryan:2014jba}. 
Thus, the possibility of the observed resonance being a part 
of an extended Higgs sector is not ruled out by the current LHC data, what 
we require is very precise measurement of various couplings of the observed Higgs boson  
to the SM particles. 

Supersymmetry (SUSY) \cite{susy,Martin:1997ns,Djouadi:2005gj} has been 
so far one of the most popular framework for 
formulating physics beyond the SM (BSM), however a SUSY signature is yet to be observed 
at the LHC \cite{cmstwiki,atlastwiki}. The Higgs sector of the Minimal 
Supersymmetric Standard Model (MSSM) is phenomenologically richer compared to that 
of the SM \cite{Martin:1997ns,Djouadi:2005gj}. The model has 
two CP-even neutral Higgses (the lighter and the
heavier ones are $h$ and $H$ respectively), one CP-odd neutral Higgs ($A$) and 
two charged scalars ($H^{\pm}$). At the tree level, the 
Higgs sector of the MSSM can be parametrised by two parameters - 
$\tan{\beta}$, the ratio of the vev of the two Higgs doublets ($H_{u,d}$), and 
the pseudoscalar Higgs mass $M_{A}$. In MSSM, the lightest CP even Higgs 
boson has a mass which lies below $M_Z$ at the tree level. To reconcile the observed Higgs mass 
at 125 GeV, one needs to invoke large higher order loop corrections 
involving the SM and SUSY particles. Soon after the Higgs discovery, 
numerous studies have been 
performed to look for additional Higgs-like states by both the ATLAS and 
CMS collaborations at the LHC. For example, both the ATLAS and CMS collaborations 
have studied the decay of the CP-even heavy scalar boson ($H$) into a pair of 
photons \cite{Aad:2014ioa,Khachatryan:2015qba}, $W$ bosons 
\cite{Aad:2015agg,Khachatryan:2015cwa}, $Z$ bosons \cite{Aad:2015kna} and 
SM-like 125 GeV Higgs bosons ($h$) 
\cite{Aad:2015uka,Khachatryan:2015yea,Aad:2014yja,Khachatryan:2016sey,Aad:2015xja} 
through various possible final state topologies. The decay of the CP-odd 
pseudoscalar Higgs boson ($A$) into a $h$ and $Z$ boson has also been 
studied at $\rm \sqrt{s} = 8~TeV$ data by both the LHC 
collaborations \cite{Aad:2015wra,Khachatryan:2015lba}. 
Moreover, the decay of the neutral Higgs bosons $H/A$ to a pair of tau-leptons has also 
been studied by the ATLAS and CMS collaboration using $\rm \sqrt{s} = 7~TeV$ 
and $\rm \sqrt{s} = 8~TeV$ data \cite{Aad:2014vgg,Khachatryan:2014wca}. This 
search mode has been found to be by far one of the most efficient channels to 
constrain the MSSM parameter space. For example, using $H/A \to \tau^+\tau^-$ 
channel, regions with large $\tan\beta$ (say $>$ 20) and small $M_A$ (say $<$ 500 GeV) 
are already excluded by LHC run-II data. Comprehensive 
studies have also been performed to search for the charged Higgs bosons. The CMS collaboration has 
looked for the charged Higgses via it's decay to $\rm \tau^{+} \nu_{\tau}$ and 
$\rm t \bar{b} $, when $\rm H^{+}$ is produced from the 
$\rm t \bar{t}$ and/or $\rm pp \rightarrow \bar{t} (b) H^{+}$ 
processes \cite{Khachatryan:2015uua,Khachatryan:2015qxa}. 
Studies of $\rm H^{+}$ decaying to a tau lepton and a neutrino has been performed 
by the ATLAS collaboration using dataset collected at $\rm \sqrt{s} = 8~TeV$ 
corresponding to an integrated luminosity of 19.5 $\rm fb^{-1}$ 
\cite{Aad:2014kga,Aad:2015typ}. 
All the above mentioned studies have not been able to find any compelling 
signature of the heavy Higgs bosons, and hence 95\% C.L. upper 
limits on the production cross-section times the branching ratios 
have been placed for a wide range of MSSM heavy resonance masses.

From the above discussions it is clear that all of the aforementioned LHC analyses of 
the MSSM neutral and charged Higgs bosons rely on their decay to the SM 
particles. However, we have to keep it in mind that the decay of these heavy MSSM Higgses 
to light supersymmetric particles, if kinematically allowed, would modify the 
branching ratios of the heavy Higgses to SM particles considerably. Let's consider 
an example: a CP even heavy Higgs $H$ can decay, if kinematically allowed, 
to a pair of neutralinos which then decays to SM Higgs and/or $Z$ and the lightest 
neutralino with $h/Z$ decaying through their usual decay modes. A few 
sample processes are shown below, 
\begin{eqnarray}
\rm {H,A} & \rightarrow & \lsptwo +  \lspone, \quad \lsptwo \rightarrow h/Z + \lspone, \nonumber \\
\rm {H,A} & \rightarrow & \lsptwo +  \lspthree, \quad \lsptwo \rightarrow h/Z + \lspone, \quad 
\lspthree \rightarrow h/Z + \lsptwo. \nonumber 
\end{eqnarray} 

Many such processes are possible, when these decays are kinematically allowed, and most 
of these decays lead to a wide range of final state topologies which can be tested at the 
LHC. Study of this very interesting possibility is the key motivation of this paper. One can 
ask the following set of questions: $(a)$ What are the relative branching ratios 
of the heavy Higgses decaying into the electroweakinos (i.e., charginos and 
neutralinos) ? $(b)$ How large/small these heavy Higgs decay modes can be satisfying the 
LHC run-I \& run-II Higgs data ? and, finally $(c)$ Do the run-II and 
future high luminosity runs of the LHC have the 
sensitivity to search for these heavy resonances through these electroweakino channels ?   
In this paper, we attempt to answer these question by performing parameter space 
scan satisfying current LHC data and then analyzing the decay of the MSSM 
heavy Higgses to electroweakinos through mono-X plus missing energy (X=$W,Z$)
and the trilepton plus missing energy signatures with a 
detailed collider study. Here we restrict ourselves within this mono-X and trilepton 
searches as these analyses are simpler and also well understood in the context of LHC. 
Moreover, note that the branching ratios of the neutralinos/charginos to $W,Z$ 
involved final states are quite large and thus we expect to have better 
sensitivity through these modes. Earlier phenomenological studies on 
the decay of $H$ and $A$ to pair of charginos and neutralinos can be seen in 
Refs.~\cite{Bisset:2007mk,Bisset:2007mi,Arhrib:2011rp,Gunion:1987ki,
Gunion:1988yc,Djouadi:1996mj,Belanger:2000tg,Bisset:2000ud,
Choi:2002zp,Charlot:2006se,Li:2013nma,Belanger:2015vwa,Ananthanarayan:2015fwa,Djouadi:2015jea}. One loop effects on 
these decay modes have also been estimated \cite{Zhang:2002fu,Ibrahim:2008rq,Heinemeyer:2015pfa}. 
From the experimental front, preliminary studies by the ATLAS and CMS collaborations 
have shown that the regions of the parameter space with heavy Higgs masses 
around 200 - 500 GeV decaying to neutralinos/charginos can be probed with 100/300 
$\ifb$ of luminosity at the 14 TeV run of LHC 
\cite{ATLAS:2009zmv,Moortgat:2001pp}. The plan of this paper can be summarized as, 

\begin{itemize}

\item We first demand that the properties of the lightest Higgs boson ($h$) 
is consistent with that of the observed 125 GeV Higgs boson satisfying 
all the present constraints from the 7 \& 8 TeV LHC data. We restrict 
ourselves within the R-parity conserving SUSY where the lightest supersymmetric 
particle (LSP), here the lightest neutralino $\lspone$, is stable 
and constitutes a source of missing transverse energy ($\met$). 

\item We then perform a random scan of the relevant MSSM parameters and 
then satisfy the updated 125 GeV Higgs data and low energy flavour data. 
After discussing salient features of the parameter space allowed by the above two 
constraints, we choose few representative benchmark points satisfying the current LHC bounds on the 
heavy Higgses as well as electroweakinos.  

\item Finally, we perform dedicated collider studies for the heavy Higgses 
associated to the selected benchmark points through mono-X 
($X = W,Z$) plus missing energy as well as trilepton 
plus missing energy ($\met$) signatures in the context of 14 TeV 
high luminosity run of the LHC. 

\end{itemize}

The rest of this paper is organized in the following order: we discuss the coupling of 
the MSSM Higgs bosons $h,H,A$ and $H^{\pm}$ to electroweakino pairs in 
Sec.\ref{sec2:couplings}. In Sec.\ref{sec2:scan} and \ref{sec2:allowed}, we discuss the various 
constraints used in our analysis and followed by the details of the 
parameter space scan. Existing collider bounds on the heavy Higgses and electroweakino 
masses are discussed in Sec.\ref{sec2:lhcbounds} followed by a discussion on some 
representative benchmark points in Sec.\ref{bp}. The collider phenomenology focusing on 
the mono-X + $\rm E_{T}^{miss}$ ($X = W/Z$) and trilepton + $\met$ 
states arising from the decay of the heavy 
Higgses to electroweakinos are discussed in Sec.\ref{monoX} and Sec.\ref{sec:3l}. 
Finally, we summarize our results in Sec.\ref{summary}.  

\section{Available parameter space and LHC data}
\label{sec2}


\subsection{Higgs couplings to Electroweakinos}
\label{sec2:couplings}

The decay of the MSSM heavy Higgs bosons to electroweakinos, namely 
charginos and neutralinos, crucially depends on the chargino/neutralino 
mixing matrices. The decay width of a generic Higgs
boson $H_k$ (with $k$=1,2,3,4 representing the $h,H,A$ and $H^{\pm}$ Higgses respectively) to pair of neutralinos 
($\lspone, \lsptwo, \lspthree, \lspfour$) and charginos ($\chonepm, \chtwopm$) 
can be written as \cite{Djouadi:2005gj}:
\begin{eqnarray}
\Gamma({H_k} \to \chii \chij)~~ \sim \quad M_{H_k} \left[\left({(g^L_{ijk})}^2 + {(g^R_{ijk})}^2\right)  
\left({1 - \frac{m^2_{\chii}}{M^2_{H_k}} - \frac{m^2_{\chij}}{M^2_{H_k}}}\right) - 4\epsilon_{i}\epsilon_{j}g^{L}_{ijk}g^{R}_{ijk} \frac{m_{\chii}m_{\chij}}{M^2_{H_k}}\right],
\end{eqnarray}
where $\epsilon_{i}$ = $\pm$1 denotes the sign of the $i$-th eigenvalue of the neutralino 
mass matrix, while $\epsilon_{i}$ = 1 the same for charginos with $\chii,\chij$ representing 
generic electroweakinos. The left- and right-handed couplings associated to the 
neutral Higgs bosons ($H_{\ell}$ = h, H, A respectively with $\ell$ = 1,2,3) with 
the neutralinos and charginos can be written as,
\begin{equation}
g_{\chii^{0} \chij^{0} H_{\ell}}^{L} = \frac{1}{2 s_{w}} \left( N_{j2} - \tan{\theta_{w}} N_{j1} \right) \left(e_{\ell} N_{i3} + d_{\ell} N_{i4}\right) + i \leftrightarrow j 
\label{HA_L_NN}
\end{equation}
\begin{equation}
g_{\chii^{0} \chij^{0} H_{\ell}}^{R} = \frac{1}{2 s_{w}} \left( N_{j2} - \tan{\theta_{w}} N_{j1} \right) \left( e_{\ell} N_{i3} + d_{\ell} N_{i4} \right) \epsilon_{\ell} + i \leftrightarrow j
\label{HA_R_NN}
\end{equation}
\begin{equation}
g_{\chii^{+} \chij^{-} H_{\ell}}^{L} = \frac{1}{\sqrt{2} s_{w}} \left( e_{\ell} V_{j1} U_{i2} - d_{\ell} V_{j2} U_{i1} \right)
\label{HA_L_CC}
\end{equation}
\begin{equation}
g_{\chii^{+} \chij^{-} H_{\ell}}^{R} = \frac{1}{\sqrt{2} s_{w}} \left( e_{\ell} V_{i1} U_{j2} - d_{\ell} V_{i2} U_{j1} \right) \epsilon_{\ell},
\label{HA_R_CC}
\end{equation}
while the same couplings for the charged Higgs bosons are given by,
\begin{equation}
g_{\chii^{\pm} \chij^{0} H^{\mp}}^{L} = \frac{\cos{\beta}}{s_{w}} \left[ N_{j4} V_{i1} + \frac{1}{\sqrt{2}} \left( N_{j2} + \tan{\theta_{w}} N_{j1} \right) V_{i2} \right] 
\label{HC_L_NC} 
\end{equation}
\begin{equation}
g_{\chii^{\pm} \chij^{0} H^{\mp}}^{R} = \frac{\sin{\beta}}{s_{w}} \left[ N_{j3} U_{i1} + \frac{1}{\sqrt{2}} \left( N_{j2} + \tan{\theta_{w}} N_{j1} \right) U_{i2} \right].  
\label{HC_R_NC}
\end{equation}

The coefficients $\rm e_{\ell}$ and $\rm d_{\ell} $ have the following definitions: 
\begin{equation}
e_{1} = \cos{\alpha}, \quad e_{2} = -\sin{\alpha}, \quad e_{3} = \sin{\beta}, \quad 
d_{1} = -\sin{\alpha}, \quad d_{2} = \cos{\alpha}, \quad d_{3} = \cos{\beta},  \nonumber 
\end{equation}
and $\epsilon_{1,2} = 1\, , \epsilon_{3} = -1$ with 
$\rm s_{w}$ = $\rm \sin{\theta_{w}}$ ($\rm \theta_{w}$ being the 
weinberg angle), $\rm \alpha$ being the Higgs mixing angle and 
$\rm \tan{\rm \beta}$ the ratio of vacuum expectation value 
of the MSSM two Higgs doublets. 

From the above-mentioned equations (\ref{HA_L_NN}) - (\ref{HC_R_NC}), one
can observe that the couplings of the MSSM Higgses with the electroweakinos 
crucially depend on their compositions. It is evident that the 
Higgses will couple with the neutralinos and charginos if and only if 
there exist admixture of the higgsinos and gauginos in the electroweakino mass 
eigenstates. The coupling of the Higgs bosons with `ino' pairs 
would be highly suppressed if both the inos are either higgsino 
dominated or gaugino dominated. Now, for example, let's consider 
the limit $\rm M_{1} < M_{2} << \mu $. From the neutralino mass matrix 
we observe that the lightest mass eigenstate $\lspone$ will be 
bino dominated and $\lsptwo$ will be wino dominated, while 
$\lspthree$ and $\lspfour$ are higgsino dominated. In this context, 
the decay modes $\rm H \rightarrow \lspone \lsptwo, \lspthree \lspfour, 
\chonep \chonem, \chtwop \chtwom$ will be suppressed, the dominant `ino' modes 
will be those which are manifestly mixtures of gaugino and higgsinos, 
like $\rm H\rightarrow \lspone\lspthree, \lsptwo\lspthree, \chonep\chtwom$ etc. 
Depending on the choice of the electroweakino mass parameters, a wide range 
of possibilities can arise which may lead to interesting phenomenological 
collider signatures. 

In order to obtain a clear picture of the above-mentioned scenario, 
in Fig.(\ref{fig:hall}) we show the branching fractions of 
the heavy Higgses $H$ (upper left), $A$ (upper right) and $H^{\pm}$ (lower) 
decaying to SM as well as modes involving electroweakinos. To generate 
the particle spectrum we use {\tt SUSPECT (version 2.43)} \cite{suspect} while 
{\tt HDECAY (version 6.41)} \cite{hdecay} has been used to calculate 
the branching fractions of the heavy Higgses. Here we vary 
$\tan\beta$ from 5 to 55 with the following choices,
\begin{eqnarray}
M_{A} = 650 ~{\rm GeV}, \quad M_{1} = 500~ {\rm GeV}, \quad M _{2} = 150~{\rm GeV}, \quad \mu = 300~{\rm GeV}, \nonumber \\
M_{3} = 5~{\rm TeV}, \quad m_{\tilde{Q}_{3l}} = m_{\tilde{t}_{R}} = m_{\tilde{b}_{R}} = 5~{\rm TeV}, \nonumber \\ 
A_{t} = -5~{\rm TeV}, \quad A_{b} = A_{\tau} = 0.
\label{couple_input_3}
\end{eqnarray}

\begin{center}
\begin{figure}[!htb]
\centering
\includegraphics[width=.4\linewidth, height=0.30\linewidth]{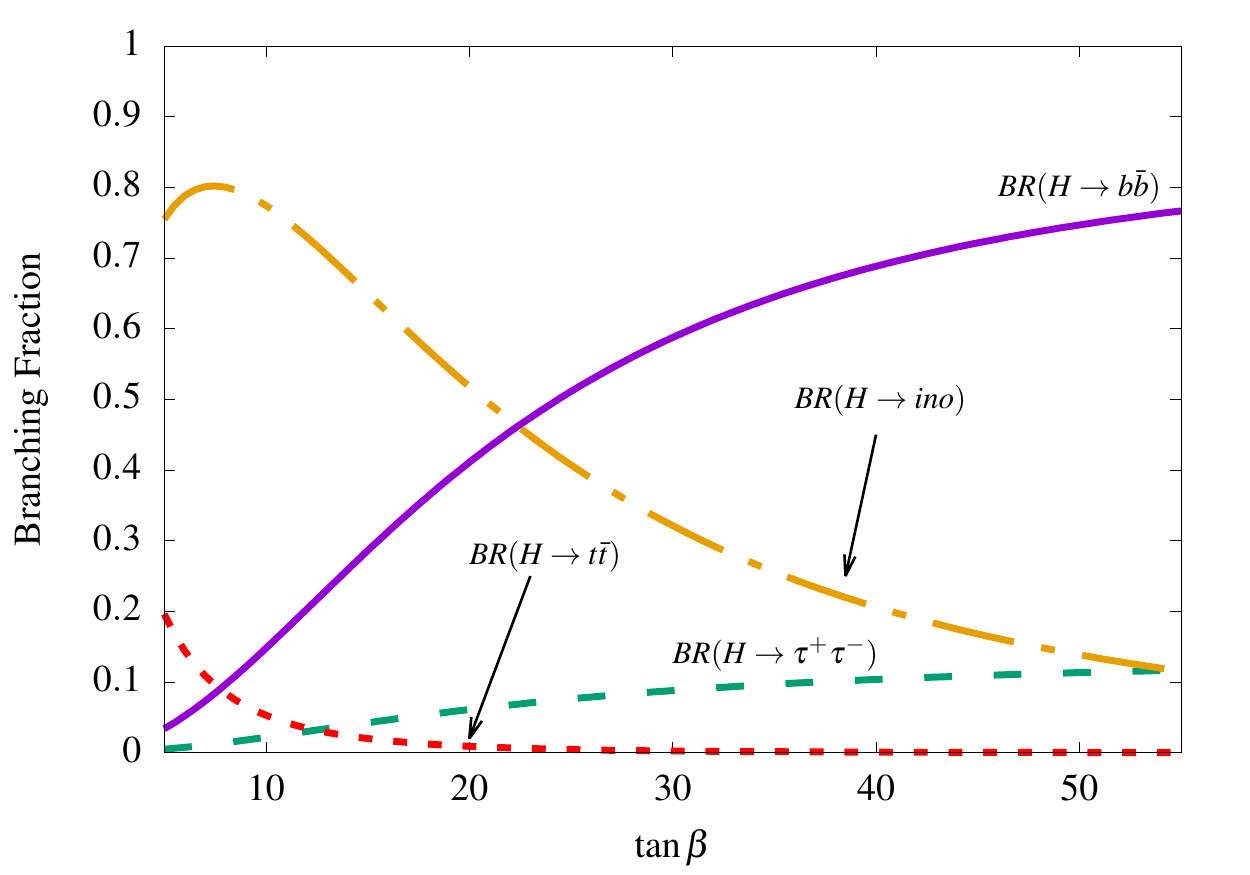}
\includegraphics[width=.4\linewidth, height=0.30\linewidth]{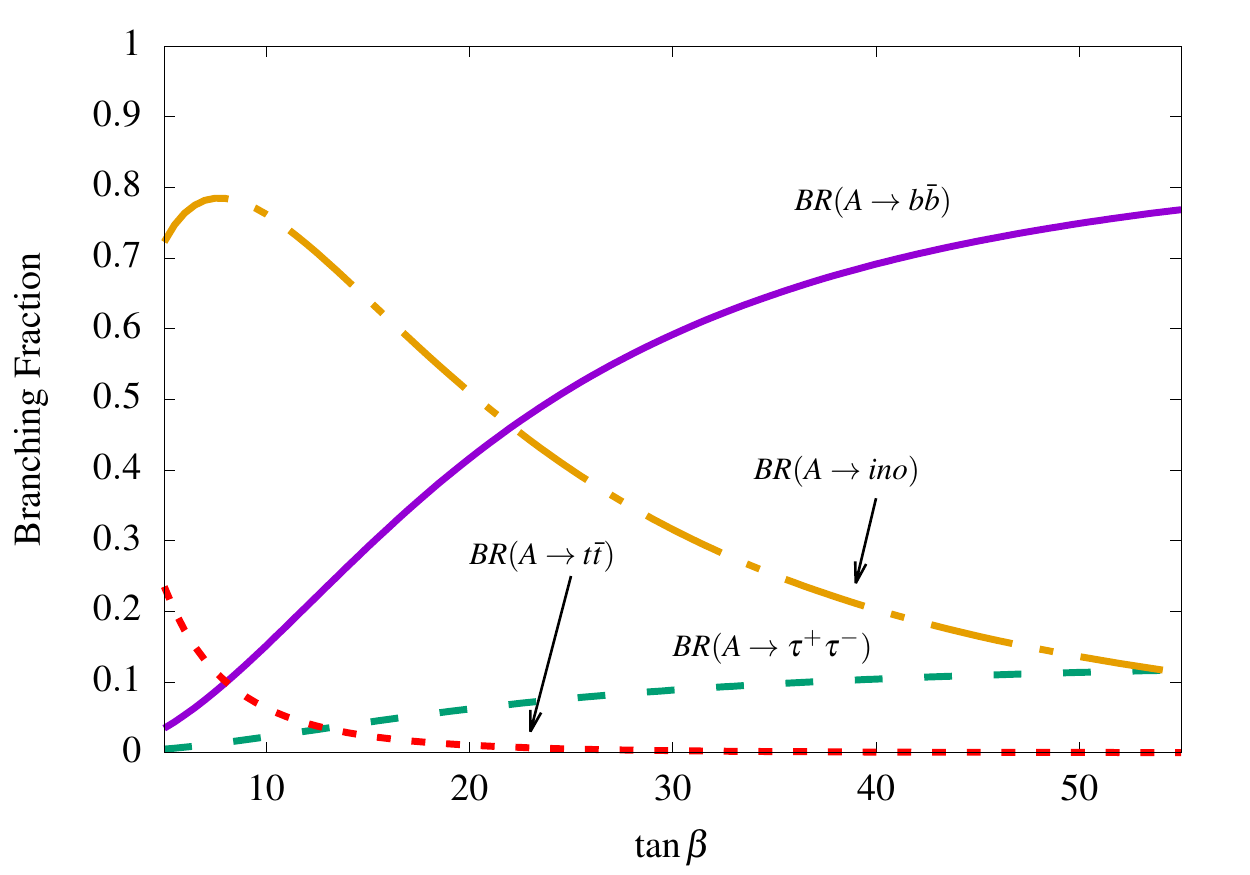} \\
\includegraphics[width=.4\linewidth, height=0.30\linewidth]{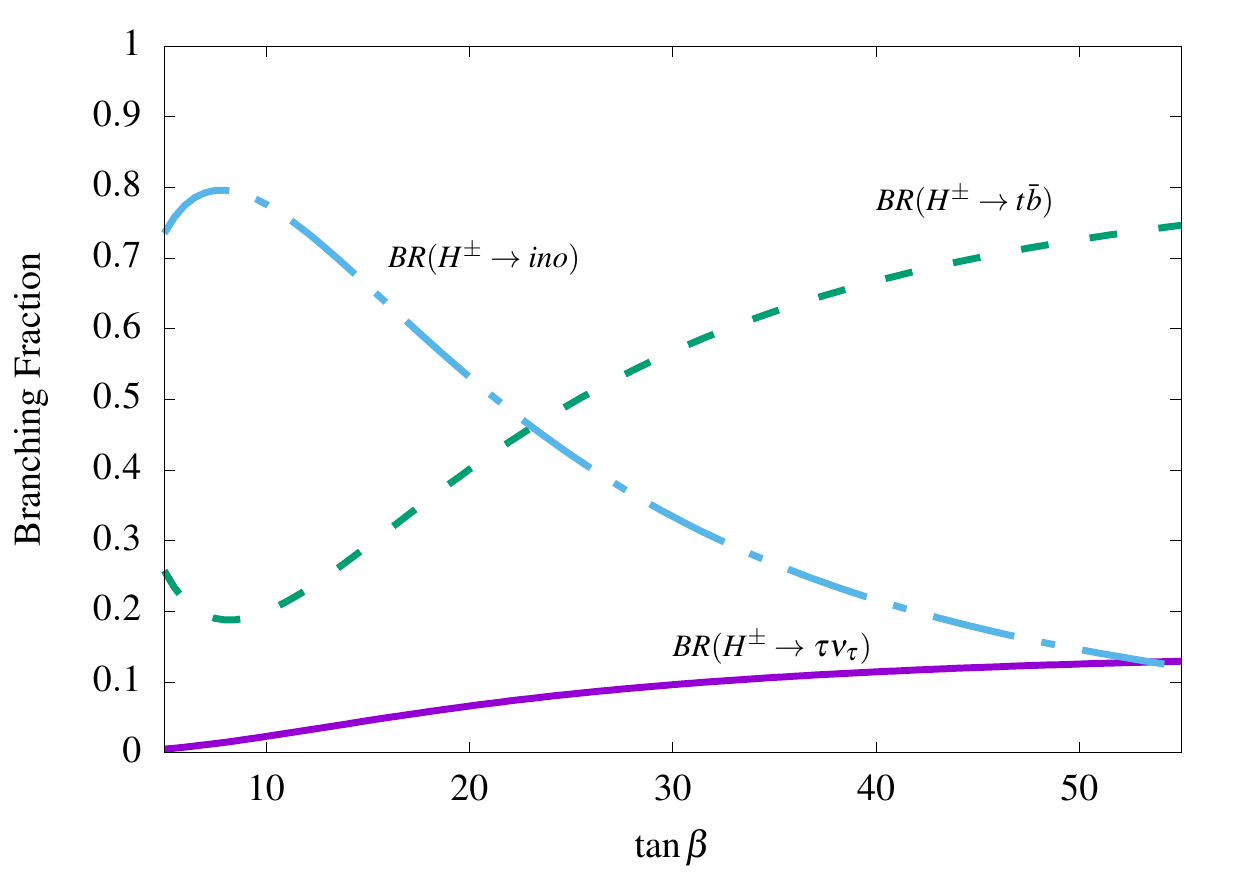}
\caption{\it The variation of the heavy Higgs ($H,A,H^{\pm}$) branching 
fractions with ${\rm \tan\beta}$ for 
both the SM and non-SM `ino' decay modes. Here the term `ino' refers to the combined chargino and 
neutralino decay modes of the respective heavy Higgs bosons.}
\label{fig:hall}
\end{figure}
\end{center}

From Fig.(\ref{fig:hall}) it is evident that we find regions\footnote{In 
this simple-minded scan, we do not impose the updated 
Higgs constraints and also any low energy physics constraints except the fact 
the lightest Higgs boson should have a mass between 122 - 128 GeV. In a more 
dedicated scan, as we discussed in next section, we do consider all these 
constraints.}
with relatively 
smaller values of $\rm \tan{\beta}$ where the non-SM decay modes 
(i.e. `ino' modes) gain dominance over the SM ones. The 
non-SM branching ratio is calculated by summing all possible `ino' decay modes of the 
Higgs boson, for example for $H$ they are $\rm H \rightarrow \lspi \lspj$ ($i,j$ = 1--4) 
and $\rm H \rightarrow \chip \chjm$ ($i,j$ = 1--2) modes. Note that, in the low $\tan{\beta}$ region, 
$\rm H \rightarrow t\bar{t}$ channel is enhanced because of $\rm \cot{\beta}$ 
proportionality factor in the $\rm Ht\bar{t}$ coupling. The 
$\rm H \rightarrow b\bar{b}$ channel dominates in the high $\tan{\beta} $ region because of 
the $\rm Hb \bar{b}$ coupling which is directly proportional to $\rm \tan{\beta}$. The behavior of the 
$\rm H \rightarrow \tau^{+} \tau^{-}$ channel is also same as the $\rm b\bar b$ mode. 
As a result of these dominant SM branching ratios, the decays of the 
neutral and charged heavy Higgses to neutralino and/or chargino pairs are suppressed at very 
low and very high $\rm \tan{\beta}$ regions. 
%
We find that, in order 
to obtain appreciable branching ratios (at least 10\% or more) to the 
non-SM `ino' modes, we are required to focus in the regions with 
moderate values of $\tan\beta$, say $\tan\beta = 5 - 20$ regime. 
\begin{center}
\begin{figure}[htb!]
\centering
\includegraphics[scale=0.50]{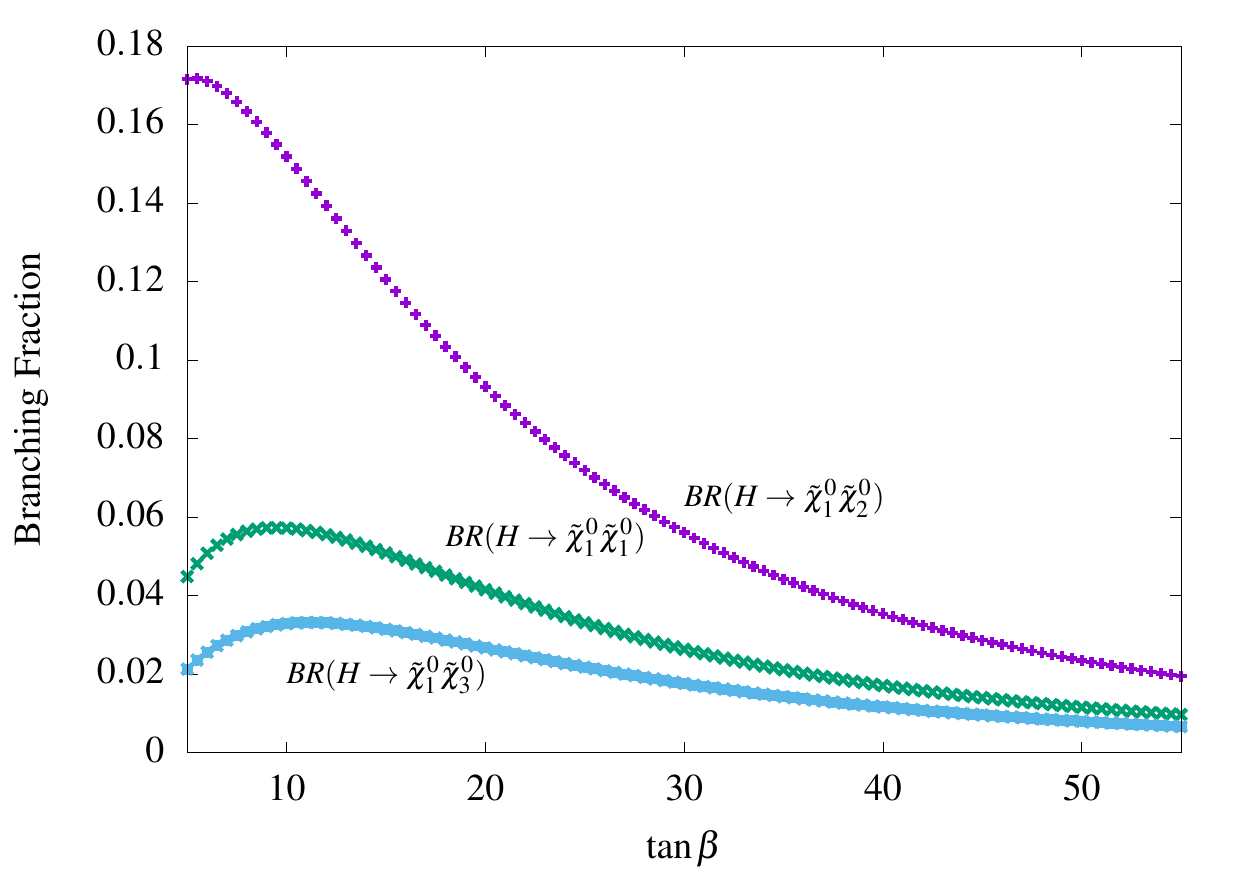}
\includegraphics[scale=0.50]{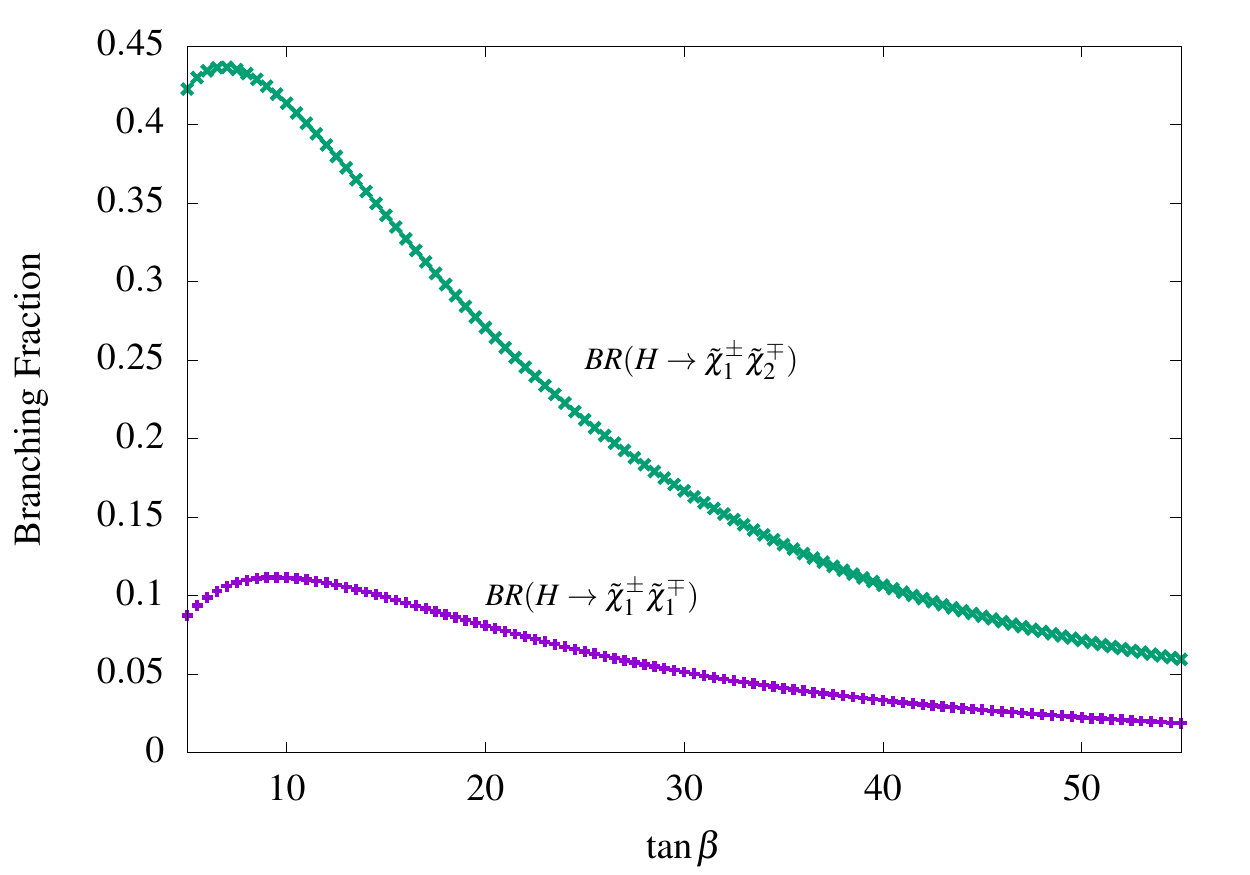} 
\includegraphics[scale=0.50]{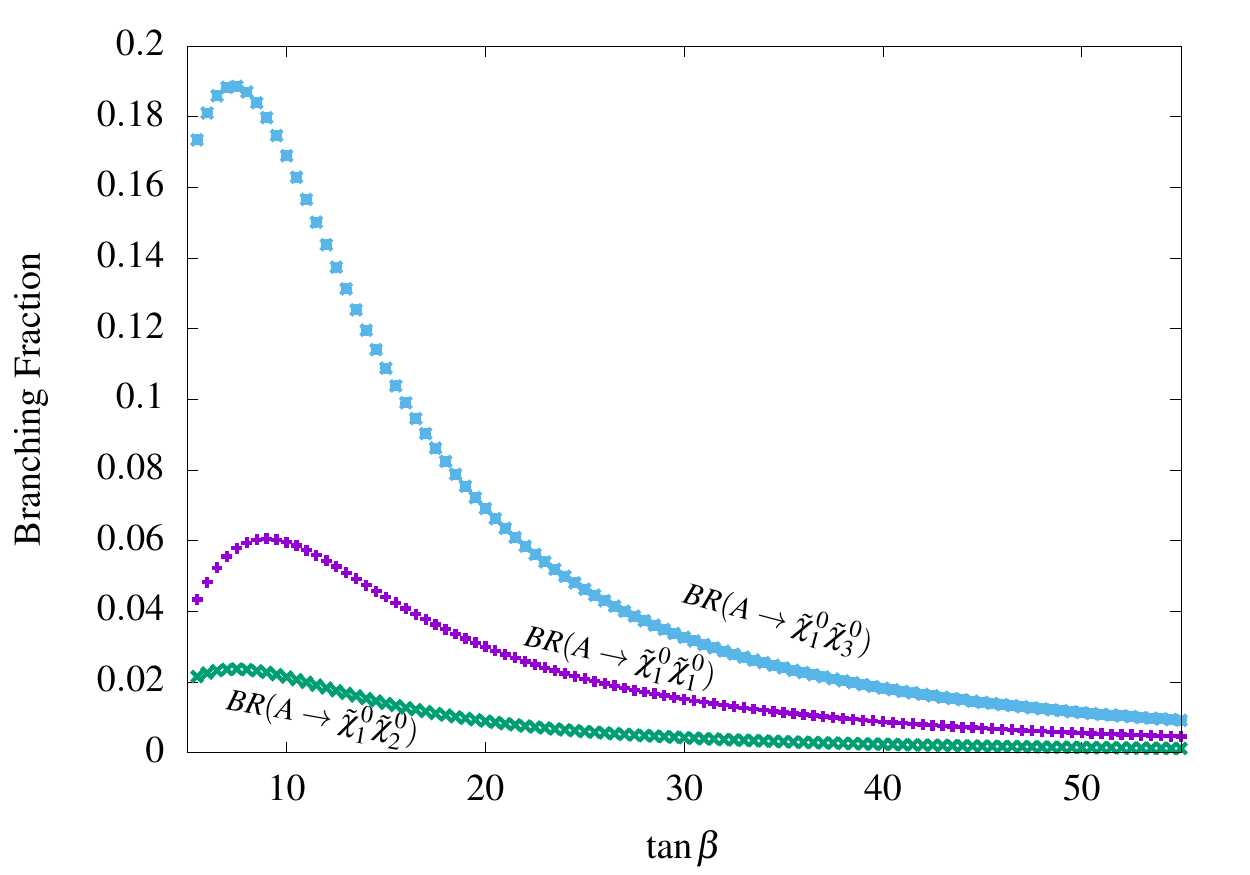}
\includegraphics[scale=0.50]{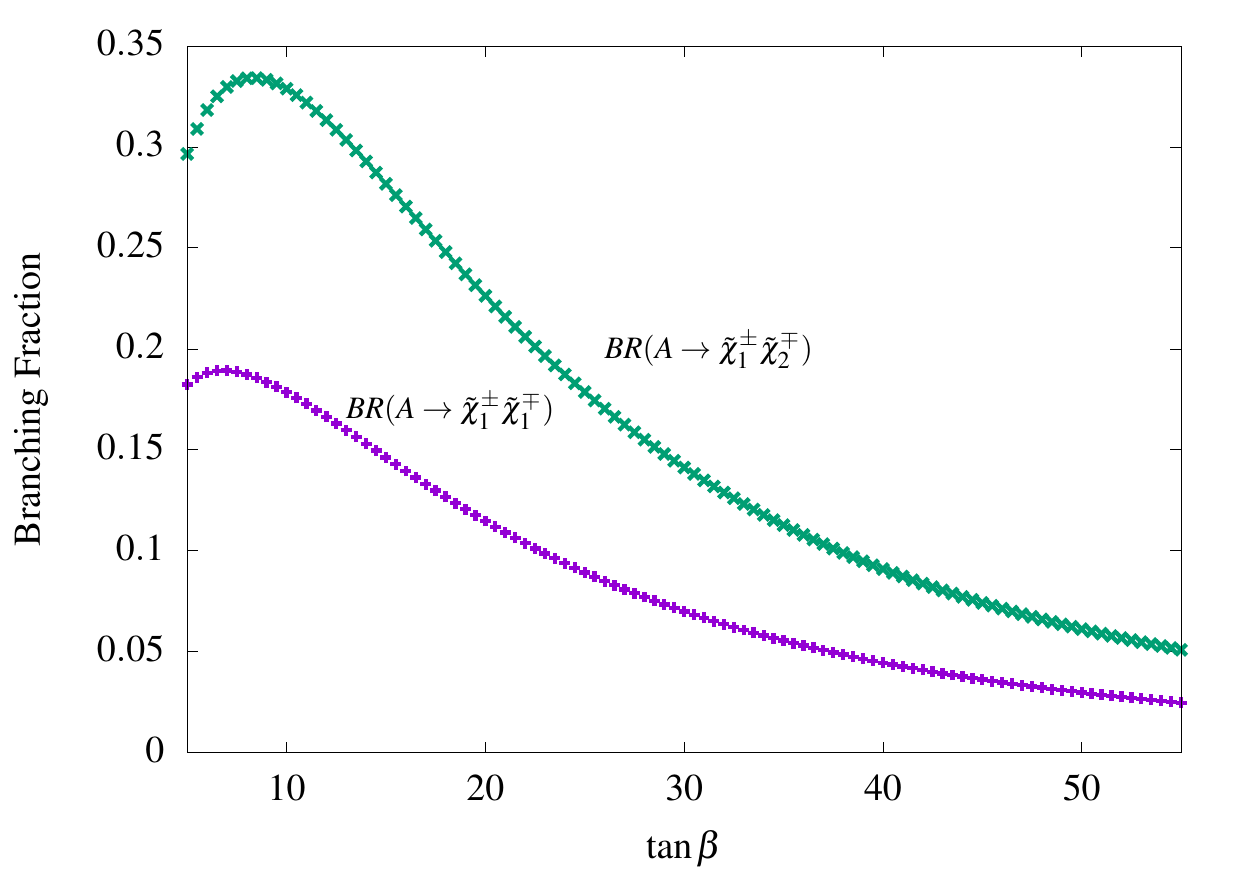}
\caption{\it The neutralino and chargino branching ratios of the heavy CP even (upper panel) and CP odd 
(lower panel) Higgs boson with respect to the variation of ${\rm \tan\beta}$.}
\label{fig:HAino}
\end{figure}
\end{center}

Before we end this section, we would like to briefly discuss how the individual `ino' 
modes behave with the variation of $\tan\beta$. In Fig.(\ref{fig:HAino}), we display 
the dominant and sub-dominant decay modes of the heavy Higgs bosons $H$ 
(upper panel) and $A$ (lower panel). The individual `ino' modes of the charged Higgs boson are 
shown in Fig.(\ref{fig:hpino}). From 
both the figures, Fig.(\ref{fig:HAino}) and Fig.(\ref{fig:hpino}), it  is evident that 
the dominant $\rm \phi \rightarrow ino$ ($\phi = H, A, H^{\pm}$) decay modes 
are those which are compounded from higgsino and gaugino mixing. For example, the 
$\rm H, A \rightarrow \chonep\chonem $ decay mode is highly suppressed as 
compared to $\rm H, A \rightarrow \chonep\chtwom$ decay mode due to small 
gaugino-higgsino mixing. Similarly, in the case of 
$\rm H,A \rightarrow \lspi\lspj$ ($i,j =$ 1--4) decay modes, those channels 
in which both the neutralinos are either higgsino dominated or gaugino dominated are 
suppressed as compared to the ones where the 
neutralinos are manifestly mixed states of gauginos and higgsinos. In summary, from 
this simple-minded scan we observe that for low to moderate values of 
$\tan\beta$, the branching ratios of the neutral Higgs $H$ and $A$ to pair 
of charginos can be as large as 50\%, while decay to pair of neutralinos can 
be around 20\%. Now, before we proceed to the next section, we would like 
to note a few important points. First, we find that these non-SM decay 
modes (i.e., Higgs decaying to electroweakinos) can be large enough, around 
30--40\%. However, so far we have not imposed the updated constraints on the 
various Higgs coupling measurements (usually expressed in terms of the signal 
strength variables) and the low energy flavour data. Moreover, there 
exists strong bounds on the masses and couplings of the heavy Higgs 
bosons from the direct searches. In addition, 
both the ATLAS and CMS collaborations at the LHC have searched for 
the electroweakinos, strong bounds exist from the LHC run-I and 
run-II data. In order to calculate the 
non-SM branching ratios 
of the heavy Higgses satisfying all the aforementioned collider constraints 
and low energy flavour data, a dedicated scan involving the MSSM parameters is required to find 
out the allowed parameter space, which is precisely the 
goal of the next section.    

\begin{center}
\begin{figure}[htb!]
\centering
\includegraphics[scale=0.5]{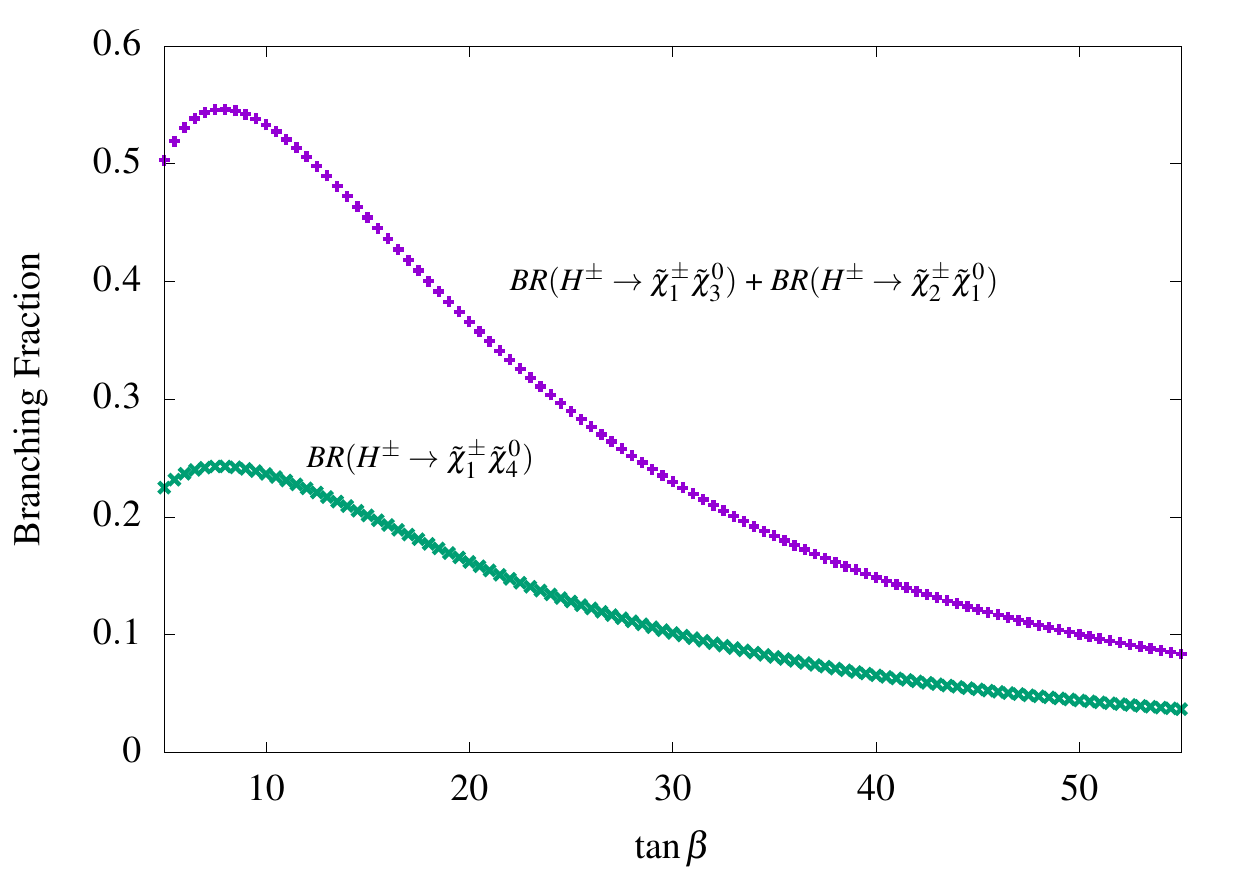}
\caption{\it The non-SM branching fractions of the charged Higgs boson 
with the variation of ${\rm \tan\beta}$.}
\label{fig:hpino}
\end{figure}
\end{center}

\subsection{Experimental inputs and Parameter space scan}
\label{sec2:scan}

In this section we describe the parameter space scan along 
with the details of various experimental constraints considered in 
our analysis. After the completion 
of LHC Run-I, both the ATLAS and CMS collaborations at the LHC have published the combined 
results of 7 and 8 TeV Higgs data 
\cite{ATLAS-CMS-comb,Aad:2015gba,Khachatryan:2014jba}. 
At the LHC, the main production mechanism of the Higgs 
boson ($h$) is the gluon-gluon fusion (ggF) process, however 
other sub-dominant production mechanisms are 
vector boson fusion (VBF), associated production with a $W/Z$ boson (Vh), 
associated production with a pair of top quarks ($t \bar t h$). Both the experimental 
collaborations have analyzed these processes for different 
decay modes of the Higgs boson like  $h \to \gamma \gamma$, $h \to WW^*$, 
$h \to ZZ^*$, $h \to b\bar b$, and $h \to \tau^+ \tau^-$ 
\cite{Aad:2015gba,Khachatryan:2014jba}. These results are usually 
presented in terms of the signal strength variables ($\mu$), defined as 
the ratio of the production cross section ($\sigma$) times the 
branching ratio ($\rm BR$) for a specific production and decay mode of the 
candidate Higgs boson in a given new physics model normalized to the SM predictions, i.e.,  
\begin{equation}
{\mu}^{f}_{i} = \frac{\sigma_{i} \times {\rm BR}^{f}}
{{(\sigma_{i})}_{\rm SM} \times {({\rm BR}^{f})}_{\rm SM}},
\label{rdef1}
\end{equation}
where $\sigma_{i}$ is the production cross-section corresponding 
to a new physics model with $i$= ggF, VBF, VH and $t\bar t H$ with 
$H$ being a generic Higgs boson and 
$f = \gamma\gamma$, $ZZ^*$, $WW^{*}$, $ b\bar b$, $\tau^+ \tau^-$ being 
the decay modes of the Higgs boson. The subscript ``SM" 
represents the respective SM expectations. These signal strength 
variables should be measured very precisely 
as any small but statistically significant deviation from the SM 
will hint at possible signatures of the beyond SM physics. 
A combined analysis using 7 and 8 TeV data performed by the ATLAS and CMS collaborations 
has derived the constraints on the individual signal strength variables as well as 
in the $\rm \mu_{ggF+tth}^{f}$, $\rm \mu_{VBF+VH}^{f}$ 
plane through a 10-parameter fit corresponding to five decay modes of the 
Higgs boson \cite{ATLAS-CMS-comb}. The above-mentioned study has assumed 
that $\rm \mu_{V}^{f}$ and $\rm \mu_{F}^{f}$ do not change with the variation of the 
center-of-mass energy from 7 to 8 TeV. In our analysis, the MSSM parameter 
space obtained after the random scan has been constrained by using the 
95$\rm \%$ C.L contours in $\rm \mu_{ggF+tth}^{f}$, $\rm \mu_{VBF+VH}^{f}$ 
plane as presented in Fig.28 of Ref.\cite{ATLAS-CMS-comb}. We study all five 
decay modes of the Higgs boson and those points which lie within the 
95\% C.L. contour are accepted for further analysis.

In addition to the updated Higgs data, we also impose 
the current flavour physics constraints on $\rm {BR (b \to s \gamma)}$ 
and $\rm {BR (B_{s} \to \mu^+ \mu^-)}$. In our analysis, we allow 
2$\sigma$ uncertainty on the measurement of these two most stringent rare 
b-decays with respect to their current measured values 
${\rm BR}(B_s \to X_{s}\gamma) = 3.43 \pm 0.22 \pm 0.21({\rm theo.})$ 
and ${\rm BR}(B_s \to \mu^+ \mu^-) = 3.1 \pm 0.7 \pm 0.31 ({\rm theo.})$ 
\cite{Amhis:2014hma}, and assume,  
\bea
 \rm 2.82\times 10 ^{-4} < BR(B_s \to X_{s}\gamma) < 4.04\times 10^{-4} \nonumber \\ 
 \rm 1.57\times 10 ^{-9} < BR(B_s \to \mu^+ \mu^-) < 4.63\times10^{-9}.  
\eea

We perform a random scan restricting ourselves to the phenomenological MSSM (pMSSM). 
The parameters that are relevant to our study are mostly associated 
to the electroweakino sector and Higgs sector of the MSSM, namely the gaugino mass 
parameters $M_{1,2,3}$, higgsino mass parameter $\mu$, pseudo-scalar 
mass parameter $M_A$, $\tan\beta$, the third generation squark trilinear couplings 
$A_t$ and $A_b$ (trilinear couplings of the sleptons and first 
two generations squarks are set to zero), third 
generation squark soft mass parameters $M_{Q_3}$, $M_{U_3}$ and $M_{D_3}$. Here we vary 
the parameters in the following ranges,
\begin{eqnarray}
\label{eq:S}
1 < \tan\beta < 55,&\ \  150~ {\rm GeV} < M_{A} < 1~ {\rm TeV}, \nonumber \\
-10~ {\rm TeV} < A_{\rm t} < 10~ {\rm TeV}, & \ \ 1~ {\rm TeV} < M_{\rm Q_3},~M_{\rm U_3},~M_{\rm D_3} < 10~ {\rm TeV}, \nonumber \\
 2~ {\rm TeV} <~M_3 ~~ < 10~ {\rm TeV} 
\label{parameterRanges}
\end{eqnarray}
while we keep the soft parameters for first two generation 
squarks and all three generation sleptons fixed at the following values,   
$ M_{\rm L_{1,2,3}} = M_{\rm E_{1,2,3}} = M_{\rm Q_{1,2}} = 
M_{\rm U_{1,2}} = M_{\rm D_{1,2}} =$ 3 TeV,      
 where $M_{\rm L_{i}}$ and 
$M_{\rm E_{i}}$ ~($i = 1,2,3$) are the left and 
right handed slepton soft SUSY breaking mass parameters, and 
$M_{\rm Q_{i}}$, $M_{\rm U_{i}}$, $M_{\rm D_{i}}$ ~($i = 1,2$) 
are the first two generation squark soft SUSY breaking mass parameters.  
The parameters  $M_1, M_2, \mu$ are varied according to some specific 
assumptions, as discussed below.

We scan over a wide range of the above mentioned parameters 
in order to obtain the lightest MSSM Higgs boson mass in the range of 
125 $\pm$ 3 GeV assuming 3 GeV uncertainty 
in Higgs mass calculation \cite{higgsuncertainty3GeV}. 
To scan the MSSM parameter space we use 
{\tt SUSPECT}
and branching ratios of the Higgs boson are evaluated using 
{\tt HDECAY}. For the estimation 
of various branching ratios of the SUSY particles (specially electroweakinos), 
we use {\tt SUSYHIT (version 1.5)} \cite{susyhit}. The flavour physics 
observables are calculated using {\tt Micromegas (version 4.1.8)} \cite{micromegas}. 
Depending on the region of our interest, we consider ten 
different type of models each differing by the values of the parameters 
associated to the electroweakino sector, namely $M_1, M_2$ and $\mu$.


\begin{table}[htb!]
\begin{center}
\begin{tabular}{|c|c|c|c|c|}
\hline
 Model Name & Parameter & Fixed & Mass & Range(s) \\
  & varied & parameters & hierarchy & of variation  \\
\hline
\hline
Model-B & $M_1$  & $M_{2} = \mu = 2~{\rm TeV} $  & $M_{2},\mu >>M_{1}$ & $1~ {\rm GeV} < M_1 < M_{A}/2$  \\[2mm]
\hline
Model-W & $M_2$  & $M_{1} = \mu = 2~{\rm TeV} $  & $M_{1},\mu >>M_{2}$ & $100~ {\rm GeV} < M_2 < M_{A}/2$  \\[2mm]
\hline
Model-H & $\mu$  & $M_{1} = M_{2} = 2~{\rm TeV} $  & $M_{1},M_{2} >>\mu$ & $100~ {\rm GeV} < \mu < M_{A}/2$  \\[2mm]
\hline
\hline
Model-BW  & $M_{1},M_{2}$  & $\mu= 2~{\rm TeV} $  & $\mu >> M_{2}>M_{1}$ & $100~ {\rm GeV} < M_{1} < M_{A}/2$  \\
        &                &                      &                       & $M_{1} < M_{2} < M_{A}$ \\ [2mm]
\hline
Model-WB & $M_{1},M_{2}$  & $\mu= 2~{\rm TeV} $  & $\mu >> M_{1}>M_{2}$ & $100~ {\rm GeV} < M_{2} < M_{A}/2$  \\
        &                &                      &                      & $M_{2} < M_{1} < M_{A}$ \\[2mm]
\hline
\end{tabular}
  \caption{\small \it  Representative models for different possible hierarchy among the 
gaugino and higgsino mass parameters. In all of these cases the pseudoscalar mass 
parameter has been varied between $150~ {\rm GeV} < M_{A} < 1000~ {\rm GeV}$. The 
`Model-X' denotes the mass parameter `X' is the lightest among all the gaugino and
the higgsino mass parameters, while `Model-XY' represents the scenario when both X and Y 
are varied independently within the specified range however with the condition $X < Y$.}
  \label{tab:model1}
\end{center}
\end{table}


Let us reiterate the point of this paper is to study 
the SUSY cascade decays of the MSSM heavy Higgses to the 
electroweakinos. So the parameters of 
interest are $M_1$ (bino), $M_2$ (wino) and $\mu$ (higgsino) mass parameters 
which are associated to the chargino and neutralino 
sector of the MSSM. Here we consider various possible 
combinations of these bino (B), wino (W) and higgsino (H) mass parameters and 
construct several representative models. From Model-B to Model-BWH 
we vary the parameters $M_{1}$, $M_2$ and $\mu$ by 
choosing all possible combinations among them (see Table~\ref{tab:model1}-\ref{tab:model2}). 
The name of each model signifies the nature of lightest electroweakino particle. For example, 
a name Model-X reflects the fact that the mass parameter `X' is the lightest among all the gaugino and 
higgsino mass parameters. Similarly, a model name `Model-XY' denotes that both X and Y are varied 
independently within the specified range, however, with the condition that X is always smaller 
than Y i.e. $X < Y$. A most general framework where no specific 
assumptions have been imposed on those parameters is labelled as Model-BWH in 
Table~\ref{tab:model2}. In all of these 
representative models, except for Model-BWH, the lightest gaugino/higgsino mass parameter is 
allowed to vary up to $M_{A}/2$, as for heavier electroweakinos the decay of the 
MSSM heavy Higgses to these modes will be kinematically forbidden. 
The lower values of some of those scanning parameters are determined from 
existing LEP bounds \cite{lepsusy}. Here, the pseudoscalar mass $M_A$ 
has been varied from 150 GeV to 1 TeV in order to obtain a wide spectrum 
of the MSSM heavy Higgs boson masses.

\begin{table}[htb!]
\begin{center}
\begin{tabular}{|c|c|c|c|c|}
\hline
 Model Name & Parameter & Fixed & Mass & Range(s) \\
  & varied & parameters & hierarchy & of variation  \\
\hline
\hline
Model-BH & $M_{1},\mu $  & $M_{2}= 2~{\rm TeV} $  & $M_{2} >> \mu > M_{1}$ & $1~ {\rm GeV} < M_{1} < M_{A}/2$  \\
        &                &                      &                          & $M_{1} < \mu < M_{A}$ \\ [2mm]
\hline
Model-HB & $M_{1},\mu $  & $M_{2}= 2~{\rm TeV} $  & $M_{2} >> M_{1} > \mu$ & $100~ {\rm GeV} < \mu < M_{A}/2$  \\
        &                &                      &                          & $\mu < M_{1} < M_{A}$ \\ [2mm]
\hline
\hline
Model-WH & $M_{2},\mu $  & $M_{1}= 2~{\rm TeV} $  & $M_{1} >> \mu> M_{2} $ & $100~ {\rm GeV} < M_{2} < M_{A}/2$  \\
        &                &                      &                          & $M_{2} < \mu < M_{A}$ \\ [2mm]
\hline
Model-HW & $M_{2},\mu $  & $M_{1}= 2~{\rm TeV} $  & $M_{1} >> M_{2} > \mu $ & $100~ {\rm GeV} < \mu < M_{A}/2$  \\
        &                &                      &                          & $\mu < M_{2} < M_{A}/2$ \\ [2mm]
\hline
\hline
Model-BWH & $M_{1},M_{2},\mu $  &        --         &             --            & $100~ {\rm GeV} < M_{2},\mu < 1000~{\rm GeV} $  \\
        &                &       --               &         --                 & $100~ {\rm GeV} < M_{1} < 1000~{\rm GeV}$ \\ [2mm]
\hline
\end{tabular}
  \caption{\small \it  Representative models for different possible hierarchy among the 
gaugino and higgsino mass parameters. In all of these cases the pseudoscalar mass 
parameter has been varied between $150~ {\rm GeV} < M_{A} < 1000~ {\rm GeV}$. A model name 
`Model-XY' represents the fact that both X and Y have been varied independently within the 
specified ranges, however, with the condition $X < Y$. The last representative model 
Model-BWH, however, represents a most general scenario when no specific mass hierarchy among 
the gaugino and higgsino mass parameters have been imposed, it just denotes the fact that 
here all the parameters are varied independently.}
  \label{tab:model2}
\end{center}
\end{table}

\subsection{SUSY decay of Heavy Higgses and the allowed parameter space}
\label{sec2:allowed}

In this section we discuss the results obtained after the parameter 
space scan for the representative models introduced in the last section 
and estimate various branching ratios of the 
heavy Higgses $H,A,H^{\pm}$ to electroweakinos. Here we would like to 
remind our readers, these branching fractions (or say 
couplings of heavy Higgses to electroweak gaugino/higgsino states) 
crucially depend on the mixing between electroweak gaugino and higgsino states 
and, in fact, a almost pure higgsino or gaugino state will have highly suppressed 
branching ratios with respect to the mixed states. This very feature has 
been observed for all of those representative models (Model-B - Model-BWH). 
For example, for Model-B, Model-W, Model-BW and Model-WB, the neutralinos and charginos are 
almost purely gaugino-like while for Model-H they are purely 
higgsino-like, and thus offering very negligible amount of mixing between the 
gaugino and higgsino states. Due to this minuscule mixing, we obtain 
highly suppressed ($\sim$ 2\%) branching fractions for all the heavy 
Higgses ($H, A, H^{\pm}$) decaying to neutralino and chargino states. 
In Table~\ref{tab:model12}, we summarize the maximum allowed branching ratios 
for the heavy Higgses summing individual decay modes of a given Higgs boson to 
different neutralino and chargino states. 

\begin{table}[htb!]
\begin{center}
\begin{tabular}{|c|c|c|c|c|c|}
\hline 
 & Model-B & Model-W & Model-H & Model-BW &  Model-WB \\ 
\hline 
$\rm H \rightarrow \lspi \lspj$ & $\rm \leq 0.07$  & $\rm \leq 0.7 $ & $\rm \leq 0.65 $ & $\rm \leq 1.2 $ & $\rm \leq 1.2 $ \\ 
\hline 
$\rm H \rightarrow \chkpm \chlmp$ & $\rm \sim 0$  & $\rm \leq 1.4 $ & $\rm \leq 1.3 $ & $\rm \leq 1.3 $ & $\rm \leq 1.4 $ \\ 
\hline 
$\rm {A} \rightarrow \lspi \lspj$ & $\rm \leq 0.07 $  & $\rm \leq 0.8 $ & $\rm \leq 0.7 $ & $\rm \leq 1.4 $ & $\rm \leq 1.4 $ \\ 
\hline 
$\rm {A} \rightarrow \chkpm \chlmp$ & $\rm \sim 0 $ & $\rm \leq 1.6 $ & $\rm \leq 1.5 $ & $\rm \leq 1.6 $ & $\rm \leq 1.6 $ \\
\hline 
$\rm H^{\pm} \rightarrow \lspi \chkpm$ & $\rm \sim 0$ & $\rm \sim 0$ & $\rm \leq 0.25 $ & $\rm \leq 1 $ & $\rm \leq 1 $ \\ 
\hline 
\end{tabular} 
  \caption{\small \it The branching ratios of the heavy Higgses (H, A and $H^{\pm}$) to 
various charginos and neutralinos. The numbers represent the sum of decay modes of the 
given Higgs boson to all possible neutralino-neutralino, neutralino-chargino and 
chargino-chargino states for models (B) to (WB) with the indices denote $i,j$ = 1,2,3,4 and $k,\ell$ =1,2.
  }
  \label{tab:model12}
\end{center}
\end{table}


\begin{figure}[!htb]
\centering
\includegraphics[scale=0.55]{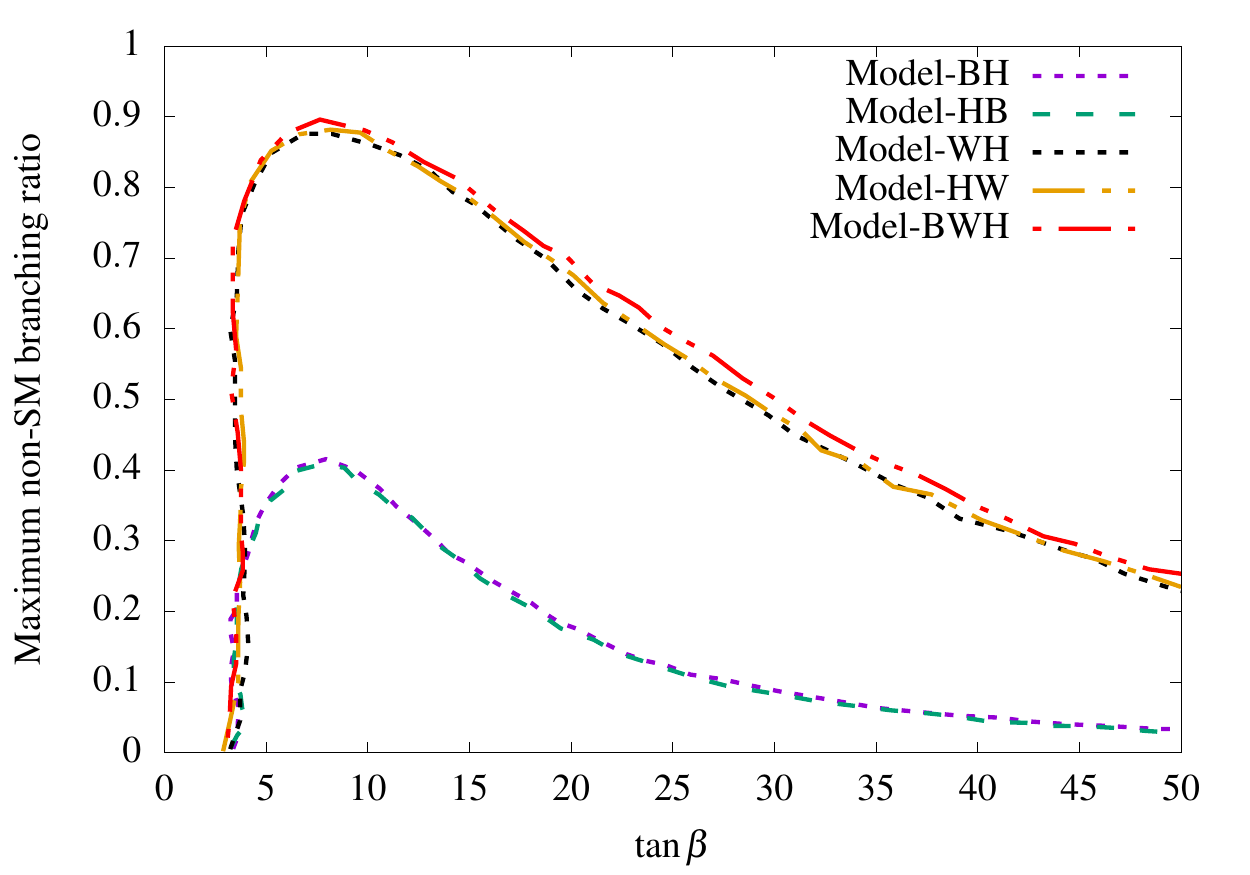}
\caption{\small \it The non-SM branching ratios for Model-BH to Model-BWH (as defined in 
Table \ref{tab:model2}) for different values of 
$\tan\beta$. Here we calculate the maximum possible value of a non-SM `ino' branching 
fraction among all possible `ino' decay modes of the heavy CP-even Higgs boson ($H$).}
\label{fig:nonSM}
\end{figure}

For Model-B to Model-WB (as defined in Table \ref{tab:model1}) by construction we have 
negligible amount of gaugino-higgsino mixing and thus the 
observation of the highly suppressed branching ratios of the 
heavy Higgses to various neutralino/chargino states is well-understood. 
Below, however, we allow significant gaugino-higgsino mixing by 
modifying the parameters and construct five more representative 
models, Model-BH to Model-BWH, and estimate the individual 
non-SM branching ratios. 
In Fig.(\ref{fig:nonSM}), we 
plot the maximum possible value of `a' non-SM branching ratio, out of 
all possible modes, obtained for a particular value of 
$\tan\beta$ in the context of all the five representative 
scenarios Model-BH to Model-BWH (as defined in
Table \ref{tab:model2}). Due to large 
mixing between higgsinos and gauginos, we find that various `ino' branching ratios 
can be as large as 40-80\%. However it is to be noted that, when both decays are 
kinematically allowed, the decay of the neutral (both CP even and odd) Higgses to a pair of 
charginos dominate over the decays to a pair of neutralinos. Moreover, when the wino-higgsino 
mixing is large compared to bino-higgsino mixing, setting $M_2$ and $\mu$ sufficiently 
light while $M_1$ decoupled, the non-SM decay modes can be significantly large. The enhancement 
is mostly driven by the large BR to the chargino pairs.

We are now in a position to discuss salient features of various representative 
models in detail. We start with Model-BH and subsequently discuss other models, 
and also compare the results. In Model-BH, 
we assume $M_2 >> \mu > M_1 $ with $M_2$ fixed at 2 TeV, and thus our choice pushes 
the masses of $\lspfour$ and $\chtwopm$ to 2 TeV. One can now expect to see several 
possible final state signatures involving lighter neutralinos and charginos, for example 
$H/A$ decaying to $\lspone \lsptwo, \lspone \lspthree, \lsptwo \lspthree$, 
$\lspone \lspone, \lsptwo \lsptwo, \lspthree \lspthree$ etc. In Fig.(\ref{fig:m6a}), we 
present branching ratios to these various final state topologies with the 
variation of $\tan\beta$. In the case when Higgs $H/A$ decays to a pair of 
neutralinos $H/A \to \lspi \lspj$ with $i \ne j$, the branching fractions 
can be as large as 20\% each, while for $i = j$ it's somewhat smaller 
(for $i=j=1$, it's around 10-15\%). Decay to a pair of charginos is also possible 
and, in fact, they can be dominant compared to the neutralino modes depending upon 
the choices of the parameters. In Fig.(\ref{fig:m6b}), we 
show various branching fractions of the $H$ (left panel) and $A$ (right panel) bosons 
to a pair of charginos and, as can be clearly seen, these branching ratios are highly 
suppressed (less than 1\%). The reason being negligibly small mixing between the 
higgsino and wino components, and thereby the lightest chargino is dominantly 
higgsino-like while heavier one is gaugino-like. The final states of the charged 
Higgs decay include chargino-neutralino pair and from Fig.(\ref{fig:m6c}) we see 
that these modes can also be sufficiently large for lower values of $\tan\beta$. 
In fact these branching fractions can go up to $\sim$ 35\% ($\chonepm \lspone$) 
and $\sim$ 20\% ($\chonepm \lspthree$) respectively, when the lightest chargino state is 
higgsino-like and the lightest neutralino is gaugino-like states. On a passing note, 
large branching ratios to these `ino' modes will lead to interesting collider signatures through 
cascade decays that can be tested at the ongoing and future runs of LHC.        
 
\begin{figure}[!htb]
\centering
\includegraphics[angle =0, width=0.45\textwidth]{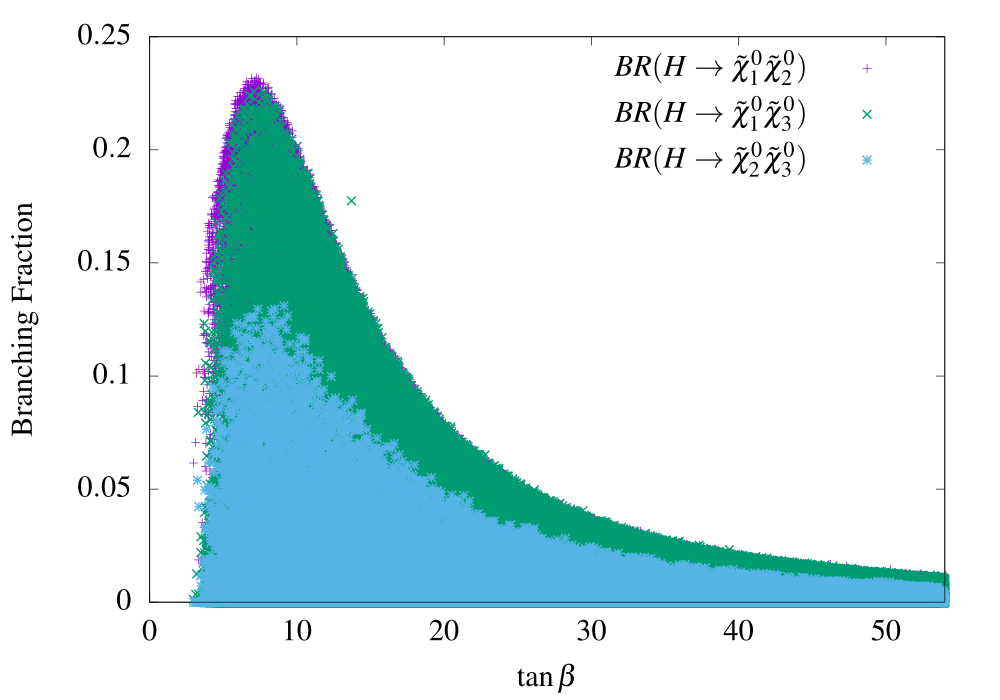}
\includegraphics[angle =0, width=0.45\textwidth]{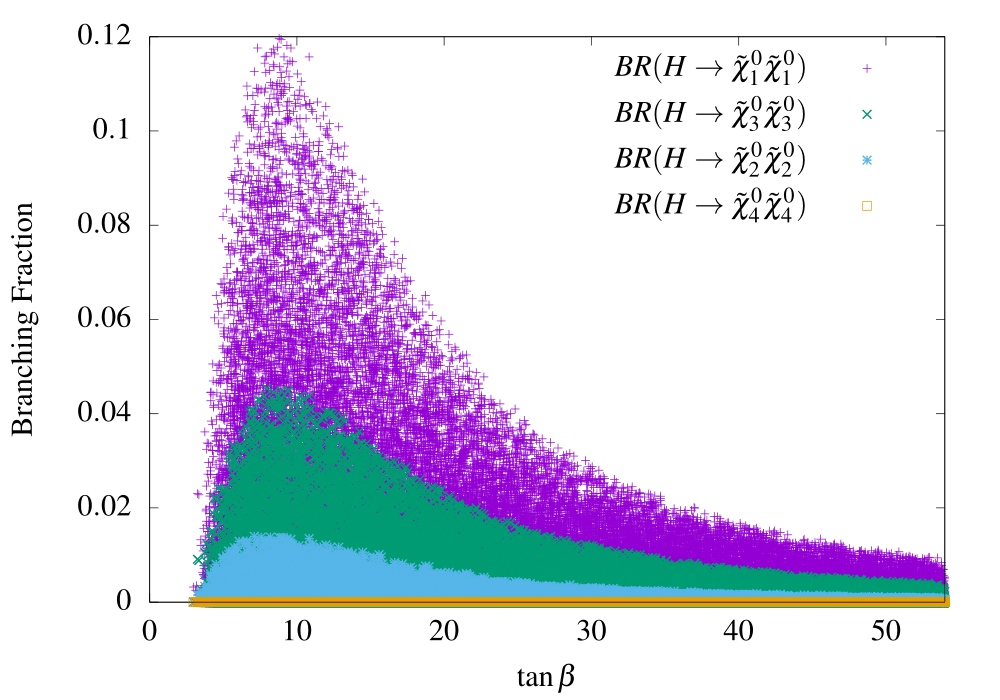}
\includegraphics[angle =0, width=0.45\textwidth]{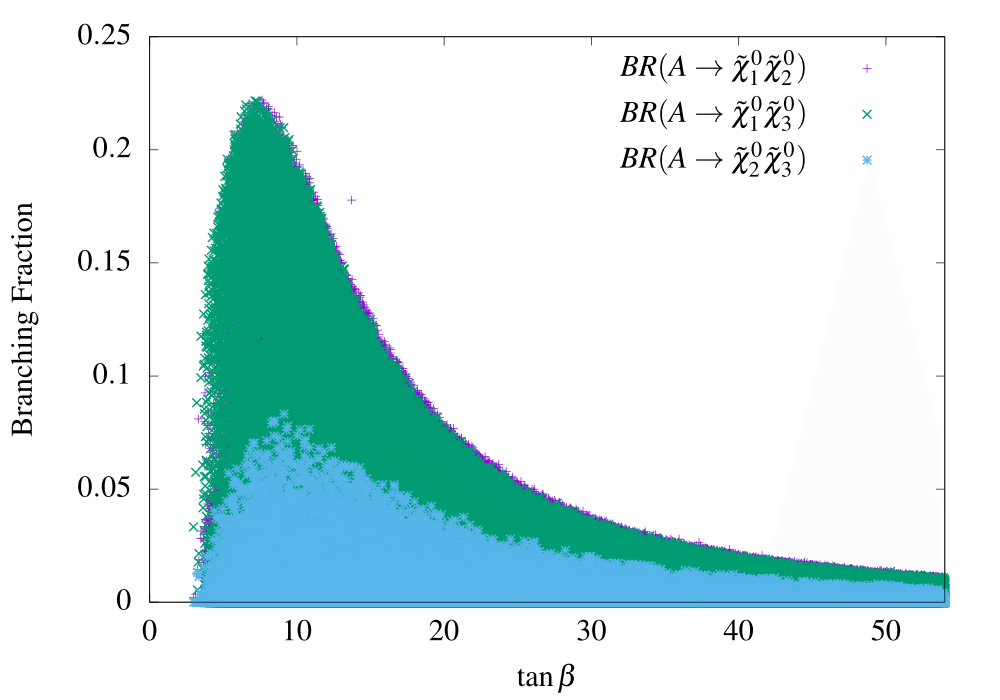}
\includegraphics[angle =0, width=0.45\textwidth]{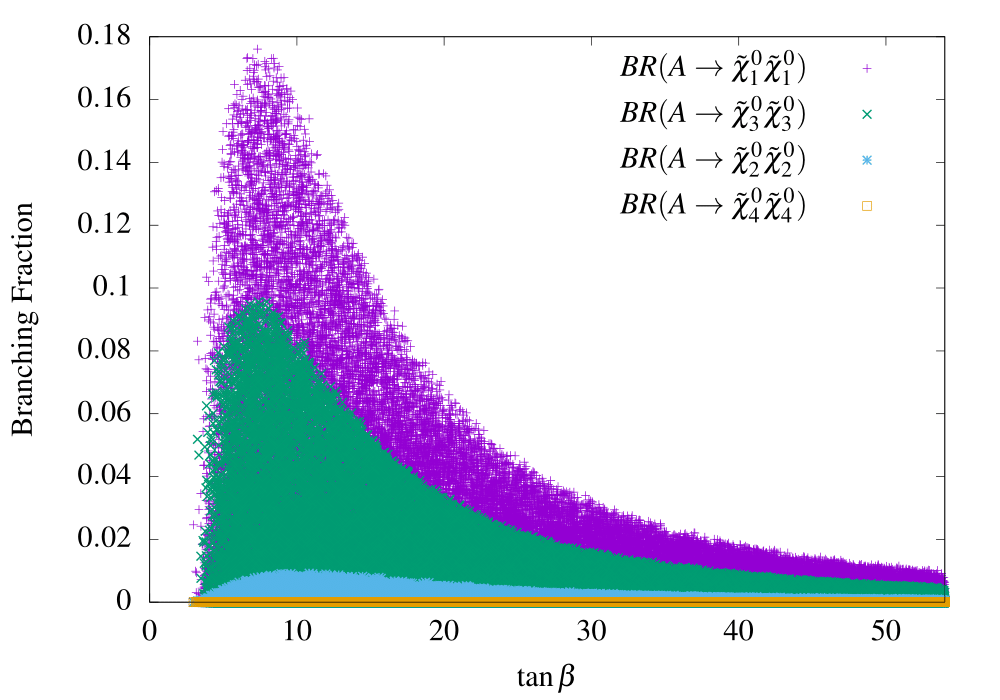}
\caption{ \small \it Branching fractions of the heavy Higgs bosons $H$ and $A$ to various 
neutralinos in the context of Model-BH.}
\label{fig:m6a}
\end{figure}

\begin{figure}[!htb]
\centering
\includegraphics[angle =0, width=0.45\textwidth]{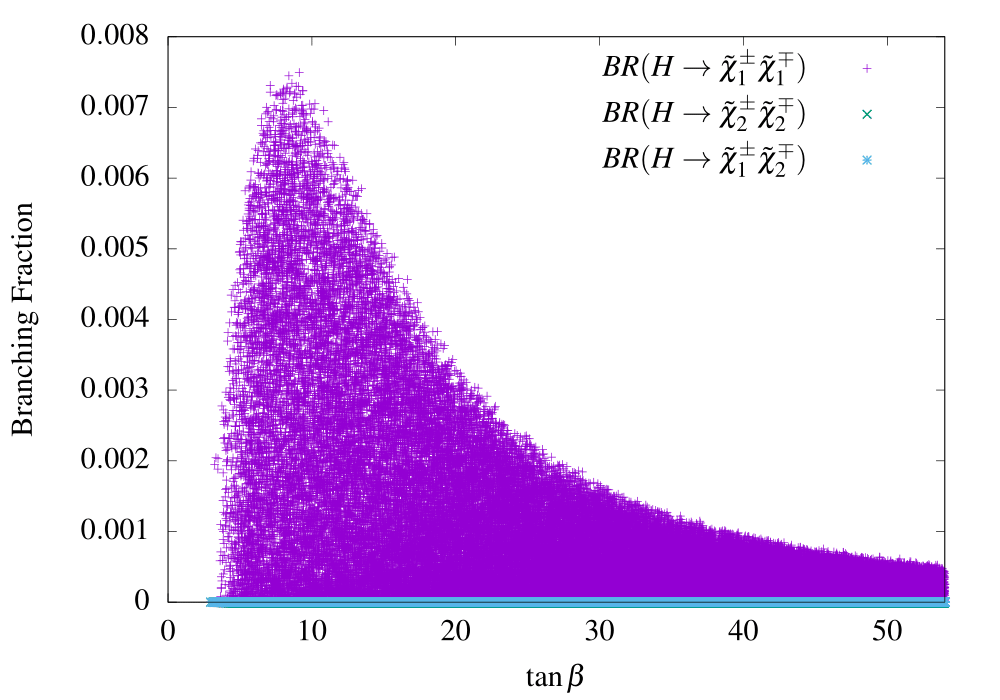}
\includegraphics[angle =0, width=0.45\textwidth]{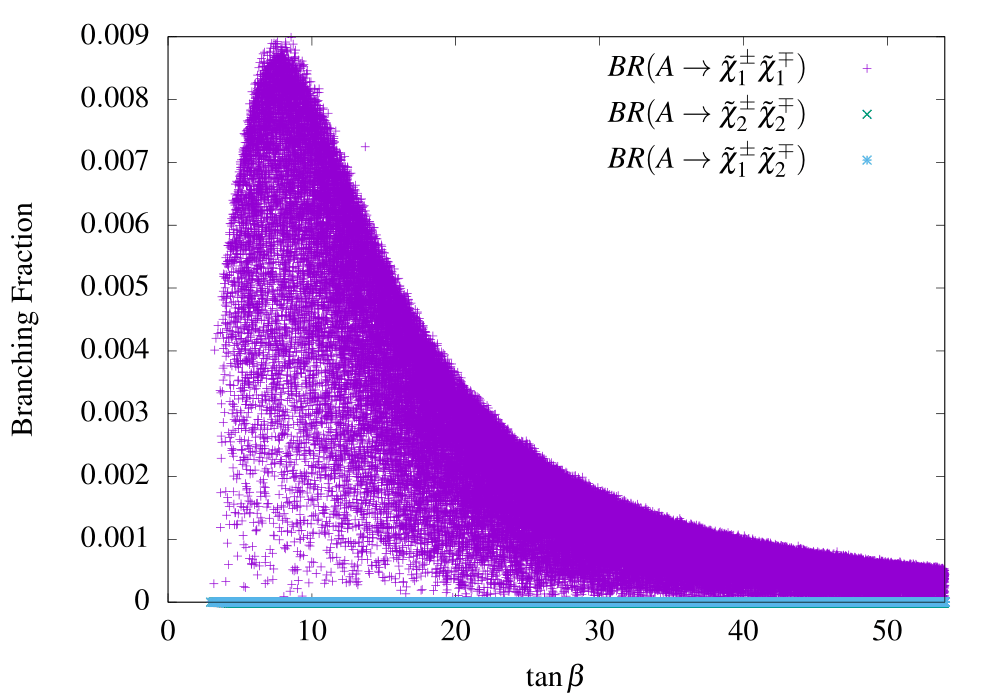}
\caption{ \small \it Branching fractions of the heavy Higgs bosons $H$ and $A$ to the 
charginos in the context of Model-BH.}
\label{fig:m6b}
\end{figure}

After having a detailed discussion for Model-BH, let is now 
interchange the hierarchy of $M_1$ and $\mu$ keeping $M_2$ 
decoupled i.e., $M_2 >> M_1 > \mu$ (Model-HB) and see how 
the results change. We find that the distributions and 
various properties are almost same as that of the Model-BH, 
except the fact that the $\lspone$ and 
$\lsptwo$ are now higgsino dominated and $\lspthree$ is now 
bino dominated (for close values of  $M_1$ and $ \mu$ 
these states are admixture of bino-higgsino). Hence compared 
to Model-BH, here the BR($H \ra$ $\lsptwo \lspthree$) and 
BR($A \ra$ $\lspone \lspthree$) dominates and 
$H/A \ra$ $\lspone \lsptwo$ has relatively small branching 
ratios. The charged Higgs branchings to electroweakinos are 
also in the same ballpark of 
model-BH. In the low $\tan\beta$ region, the maximum BRs of 
$H^\pm$ are about 35\% for $\chonepm \lspthree$ and 20\% 
for $\chonepm \lspone$ respectively.

One can also assume a possible hierarchy between the model parameters 
as $M_1 >> \mu > M_2$ (Model-WH) or, $M_1 >> M_2 > \mu$ (Model-HW). 
The difference between these two scenarios are as follows: in the first 
case LSP is wino dominated (Model-WH), while in the second case 
LSP is higgsino dominated (Model-HW). We find that, depending on the 
wino-higgsino mixing and the values of $\tan\beta$, heavy 
Higgs decaying to a pair of neutralinos and/or charginos can reach up to 10 - 25\%. However, 
due to the presence of light charginos in the particle spectrum and 
large higgsino-wino mixing, heavy Higgs decaying to a pair of charginos can be as large as 50\% for 
both Model-WH and Model-HW. Moreover, we also find that the charged 
Higgs decay to a chargino-neutralino pair is also possible (with BR $\sim$ 25\%) in 
the entire region of the parameter space for low to moderate values of 
$\tan\beta$.  


\begin{figure}[!htb]
\centering
\includegraphics[scale=0.25]{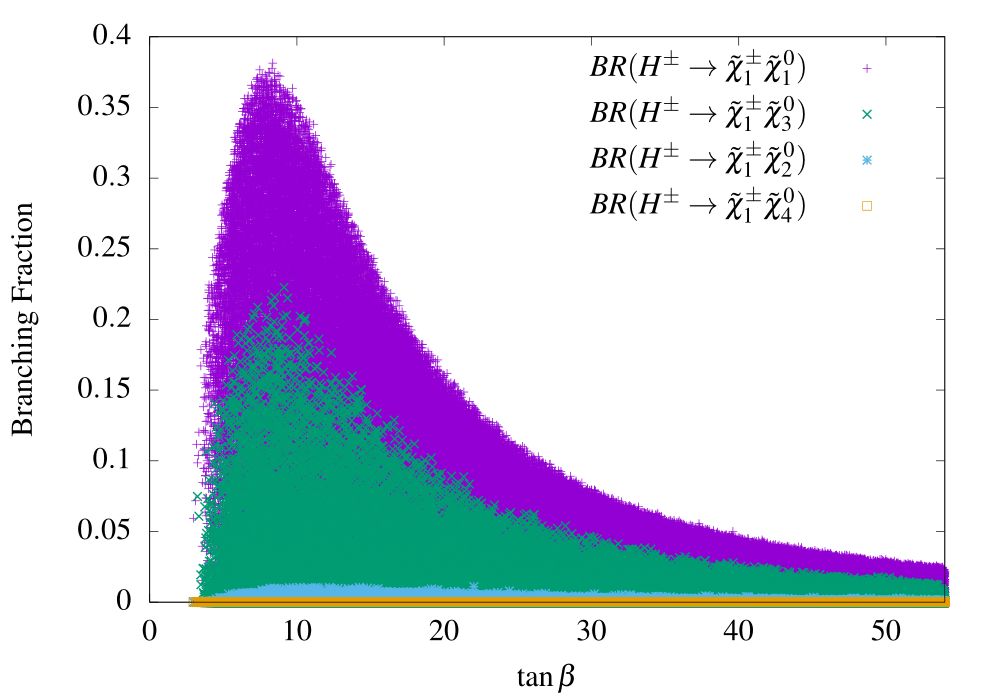}
\caption{ \small \it The decay of the charged Higgs boson $H^{\pm}$ to chargino-neutralino 
final states, where we consider Model-BH.}
\label{fig:m6c}
\end{figure}

\begin{figure}[!htb]
\centering
\includegraphics[angle =0, width=0.42\textwidth]{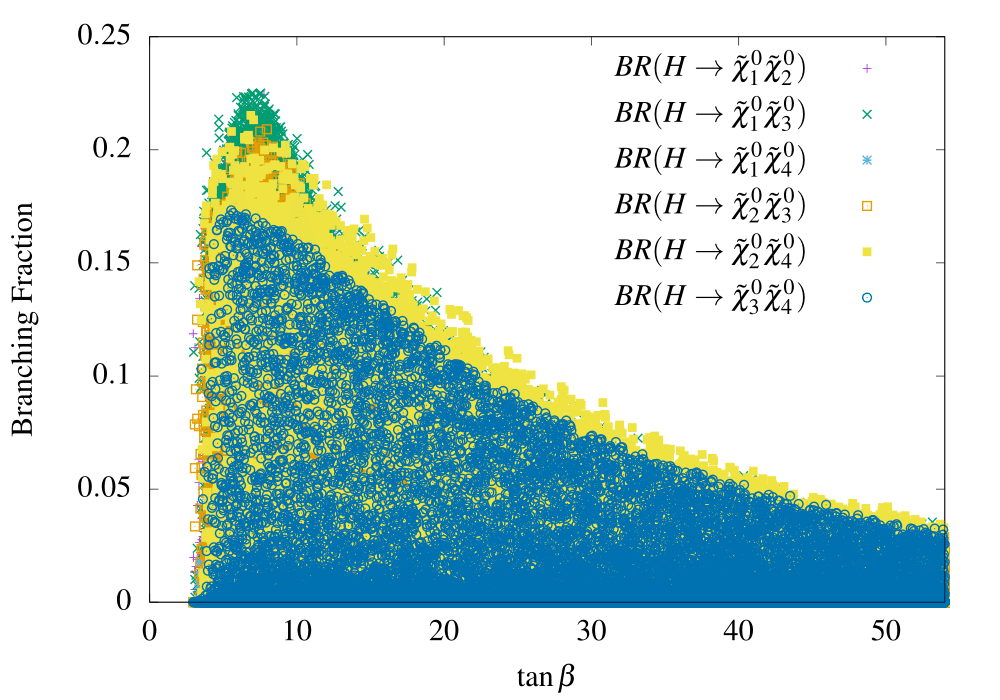}
\includegraphics[angle =0, width=0.42\textwidth]{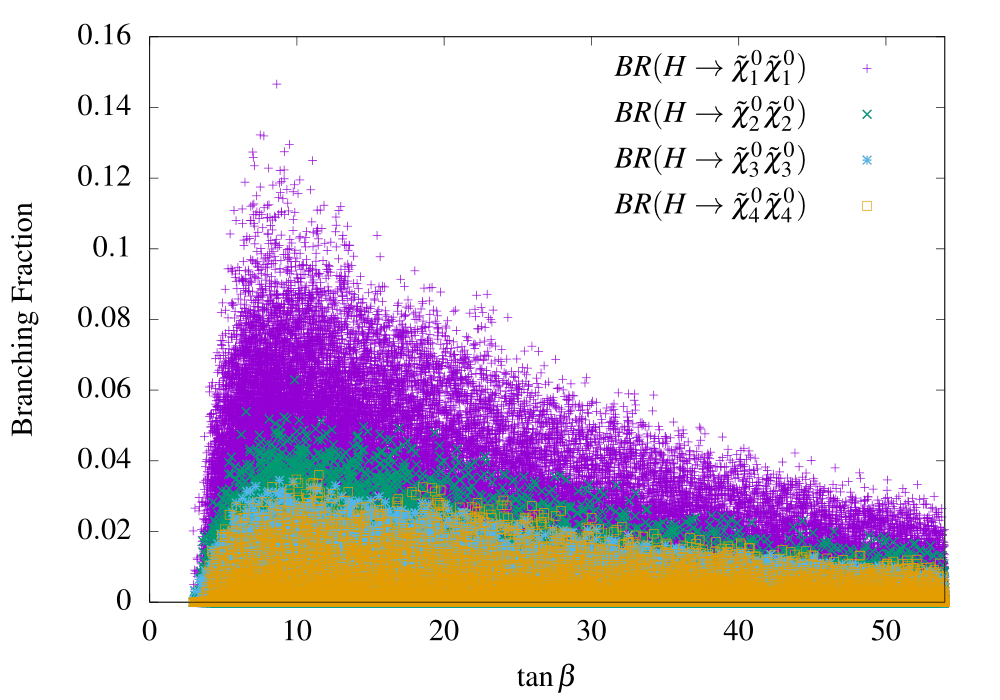}
\includegraphics[angle =0, width=0.42\textwidth]{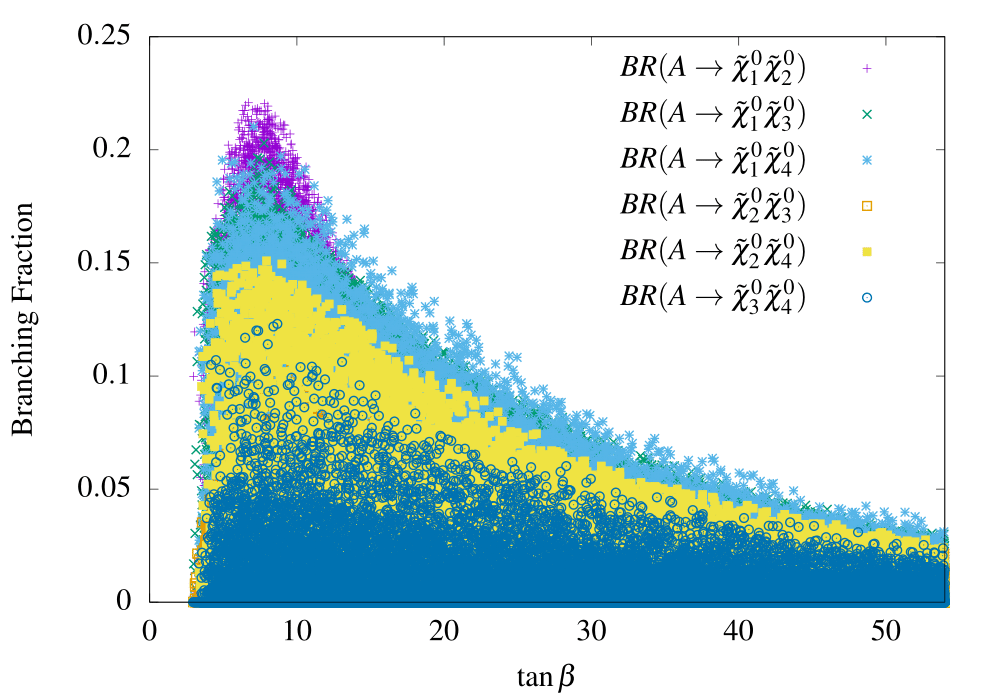}
\includegraphics[angle =0, width=0.42\textwidth]{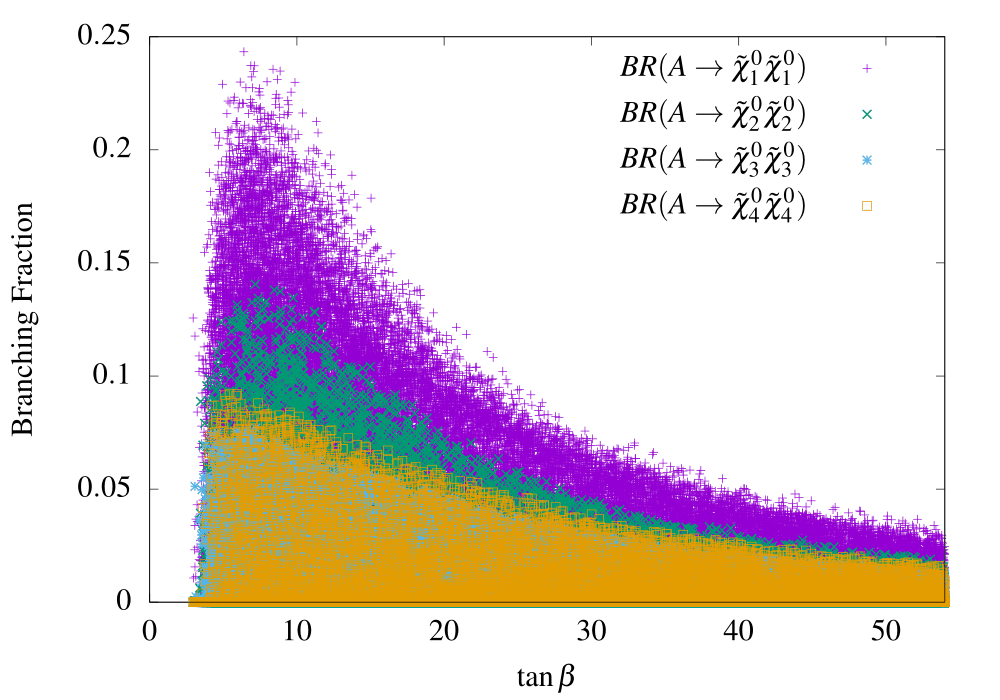}
\includegraphics[angle =0, width=0.42\textwidth]{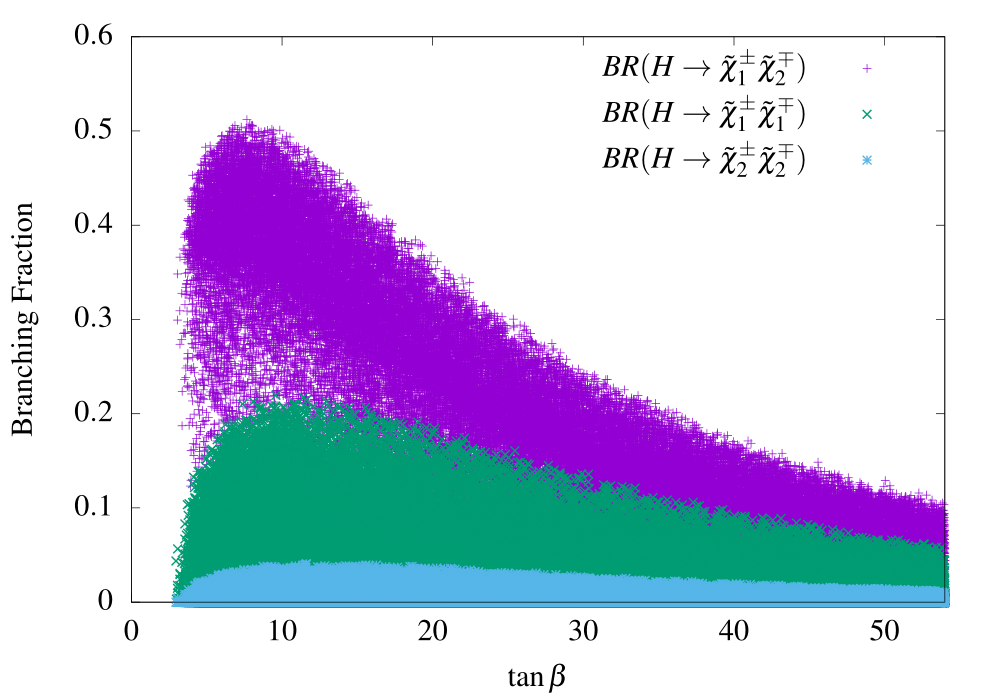}
\includegraphics[angle =0, width=0.42\textwidth]{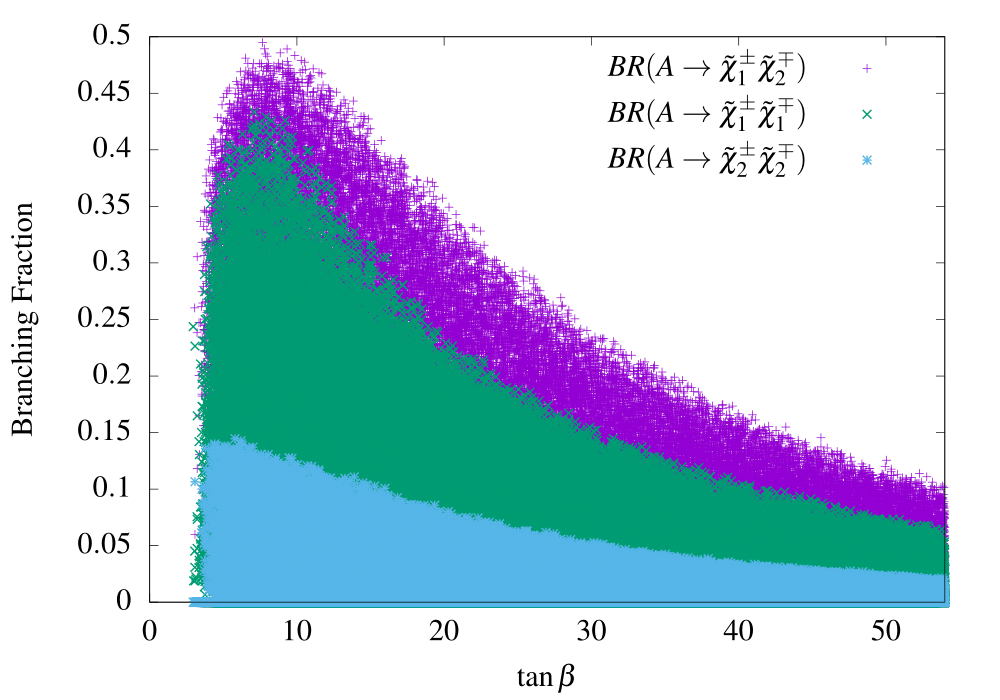}
\includegraphics[angle =0, width=0.42\textwidth]{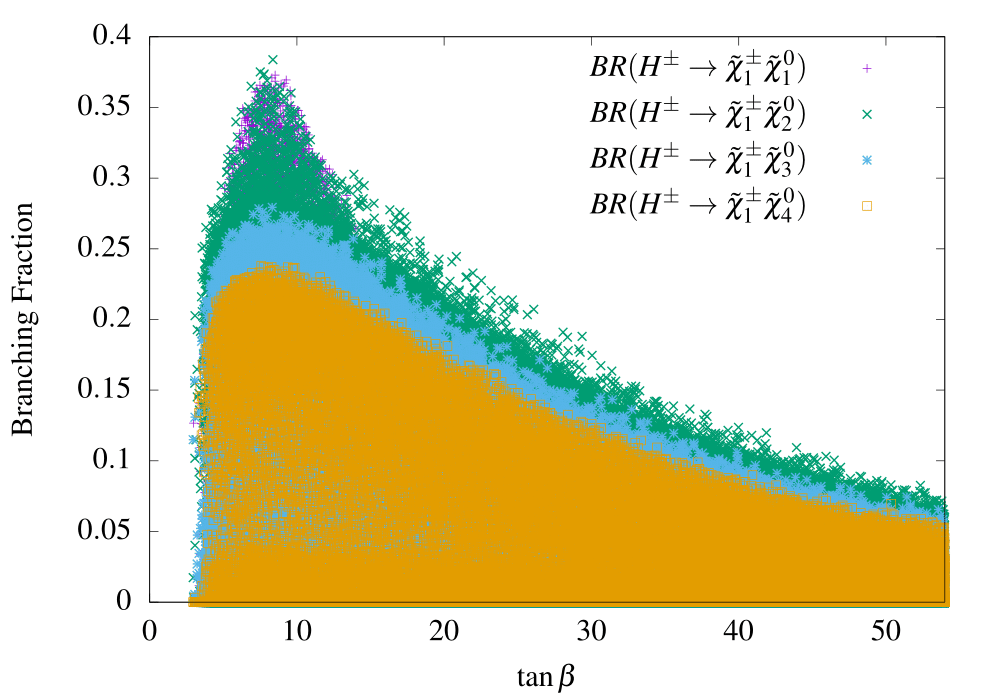}
\includegraphics[angle =0, width=0.42\textwidth]{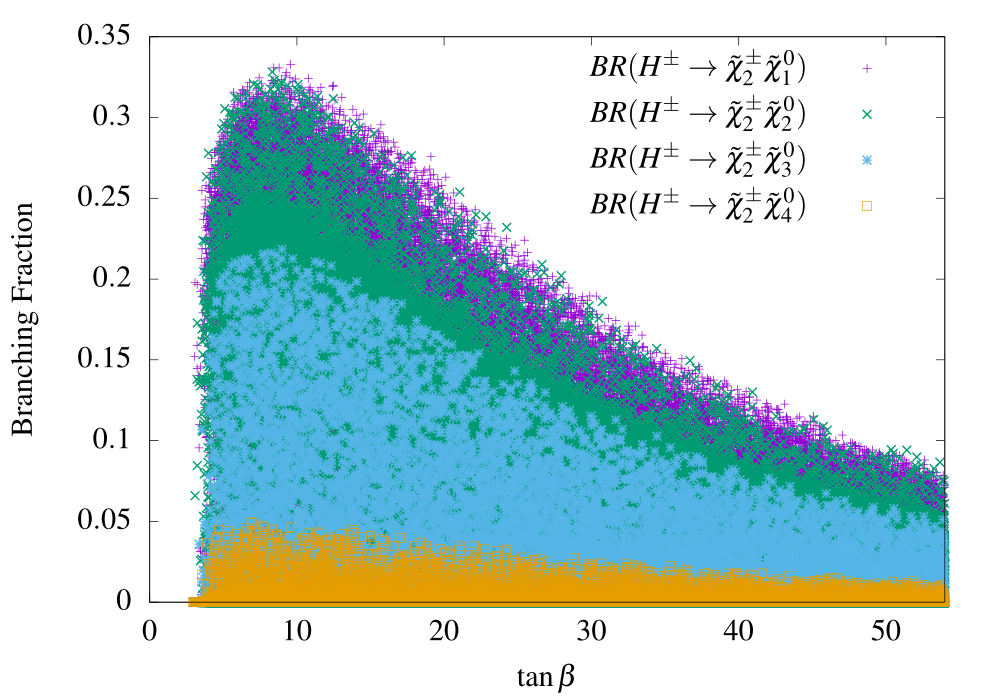}
\caption{\it Branching ratios for heavy Higgses (neutral and charged) decaying 
to various chargino and neutralino final states in the context of Model-BWH.}
\label{fig:m10}
\end{figure}

Given the three free input parameters, $M_1$, $M_2$ and $\mu$, one is not 
always forced to assume a specific hierarchy between them, instead one can 
always take a simple-minded approach allowing all of them to vary independently. This 
is precisely our last representative model, namely Model-BWH where we scan for all 
possible values of $M_1$, $M_2$ and $\mu$ without imposing any hierarchy 
among them. The added advantage of this kind of choice is many decay modes, 
which were not kinematically allowed earlier, now will open up and thus will make it 
phenomenologically richer. For example, the presence of small $M_1$, $M_2$ and 
$\mu$ implies that the decay modes containing $\lspfour$ are now accessible in 
some parts of the parameter space. Again, however, magnitude of the individual 
decay modes will crucially depend upon the values of $M_1$, $M_2$ and $\mu$. 
In Fig.(\ref{fig:m10}), we display all the possible decay modes of the heavy 
neutral and charged Higgs bosons to final states with charginos and neutralinos 
for Model-BWH. From the 
distributions one can observe appreciable amount of enhancement 
in heavy Higgs branching fractions for some of the decay modes which 
were absent in our earlier constructions (say Model-B to Model-WB). Interestingly, 
in all of these cases, long cascade decays originating from heavy Higgses decay 
may result into testable collider signatures, starting from relatively simpler signatures 
like mono-$Z$/mono-$W$/mono-$h$/dibosons plus $\met$ signatures, 
to relatively complex final state topologies consisting of multiple leptons, 
jets and $\met$.   

%
%

For rest of our analysis, we consider 
those benchmark models where these heavy Higgses have 
appreciable amount of branching fractions to charginos and neutralinos, and 
thus we restrict ourselves within Model-BH to Model-BWH. However, before we 
advance towards the detailed collider analysis, we would like to 
remind our readers that there exists avalanche of studies where 
both the ATLAS and CMS collaborations at the LHC have searched for heavy 
scalar resonances through different final state signatures. Moreover, 
direct searches for the electroweakinos have also been performed at the LHC. 
However, it is difficult to incorporate all these LHC constraints into 
our parameter space scan. Thus, in order to enliven our analysis, 
we choose few representative benchmark points which are allowed by 
the current LHC data and then perform the detailed collider analysis.


\subsection{Current LHC bounds for heavy Higgs sector and electroweakinos}
\label{sec2:lhcbounds}

In this section we summarize the relevant LHC constraints originating mainly 
from the LHC run-I data and also a few from the early 13 TeV run-II data. 
We start by discussing the current bounds on the masses and branching ratios 
of the heavy neutral and charged Higgs bosons obtained by the 
ATLAS and CMS collaborations for various possible final states, and then 
proceed to outline the present constraints on the electroweakinos from the run-I data.     

\subsubsection{Search for $H$ with $\gamma\gamma$ final states}
\label{sec:Hgamgam_8}

Di-photon invariant mass distribution can be a useful probe to 
search for the additional Higgses. Both ATLAS and CMS collaboration 
have looked for scalar particles decaying via narrow resonances 
into two photon final state using the run-I data 
\cite{Aad:2014ioa,Khachatryan:2015qba}. The ATLAS (CMS) 
collaboration has performed the study using 20.3 (19.7) $\ifb$ 
data in the mass range 65 - 600 (150 - 850) GeV and no significant 
excess over the SM background has been observed. The results have been 
presented as 95$\rm \%$ C.L upper limits on the production cross-section 
times branching ratio into pair of photons i.e., limit on the quantity 
$\sigma \times {\rm Br}(H \to \gamma \gamma)$. For example, the 
upper limit\footnote{Note that, ATLAS has obtained upper limit on fiducial cross-section ($\sigma_{fid}$)
times branching ratio, where $\sigma_{fid}$ includes an efficiency correction factor
\cite{Aad:2014ioa}.} on $\sigma \times {\rm Br}(H \to \gamma \gamma)$ 
for a heavy Higgs with mass $M_H$ = 600 GeV as obtained by the ATLAS is 
around 1 ${\rm fb}$ \cite{Aad:2014ioa}.

Recently, both the ATLAS and CMS collaborations have reported an 
excess in the di-photon invariant mass distribution around 750 GeV with 
the early 13 TeV data \cite{Aaboud:2016tru, Khachatryan:2016hje}. 
The local significance of the excess observed by CMS is approximately 
3.4$\sigma$ combining 8 TeV 19.7 $\ifb$ data and 13 TeV 
3.2 $\ifb$ data \cite{Khachatryan:2016hje}. ATLAS has also observed 
the excess with local significance 3.9$\sigma$ for about 50 GeV decay width 
of the heavy resonance \cite{Aaboud:2016tru}. In our work, we use the 
95\% C.L upper limits on the production cross section times branching ratio 
for different choices of heavy resonance masses as derived by both the 
experimental collaborations.

\subsubsection{Search for $H$ with $WW$ final states}
\label{sec:HWW_8}

Search for a high-mass Higgs boson in the 
$H \to WW \to \ell \nu \ell \nu$ and $H \to WW \to \ell \nu q q$ 
decay channels has been performed by the ATLAS and CMS collaborations 
using the LHC run-I data \cite{Aad:2015agg, Khachatryan:2015cwa}. The 
absence of an evidence of a signal, upper limits on 
$\sigma \times {\rm Br (H \to WW)}$ as a function of $m_H$ 
are reported for different possible choices of the width ($\Gamma_H$) of the 
Higgs boson. The ATLAS study has been performed using 
20.3 $\ifb$ data in the mass range 200 - 1500 GeV 
\cite{Aad:2015agg} with 8 TeV data. CMS has also performed 
the same analysis using the complete data set 
(5.1 + 19.7 $\ifb$) collected at $\sqrt s$ = 7 \& 8 TeV data 
in the mass range 145 $< m_H <$ 1000 GeV \cite{Khachatryan:2015cwa}. 
Here we would like to note that as we proceed near the alignment 
limit \cite{Gunion:2002zf} 
(i.e. $(\beta - \alpha) \sim \frac{\pi}{2}$ with $\alpha$ being the Higgs 
mixing angle), BR($H \to WW)$ become highly suppressed and then bound on 
$\sigma \times {\rm Br}(H \to WW)$ become less significant for most of 
the MSSM parameter space (see Fig 7b of Ref \cite{Bhattacherjee:2015sga} for more details). 
The ATLAS collaboration has also looked for a heavy neutral scalar particle 
decaying into pair of $W$ bosons, and placed upper limits on the 
$\sigma \times {\rm Br}(H \to WW)$ as a function of $m_H$ in the entire 
mass range of 600 GeV to 3 TeV using the LHC run-II data with 
$\lum = 3.2~\ifb$ luminosity \cite{atlas_H_WW_13tev}.


\subsubsection{Search for $H$ with $ZZ$ final states}
\label{sec:HZZ_8}

Searches for additional Higgses in the $ZZ$ channel have also been 
reported by the ATLAS and CMS collaborations 
\cite{Aad:2015kna, Khachatryan:2015cwa, atlas_H_ZZ_13tev}. The analysis 
by the ATLAS with 20.3 $\ifb$ of 8 TeV data focuses on the four final 
state topologies: $H \to ZZ \to \ell^+ \ell^- \ell^+ \ell^-, 
\ell^+ \ell^-\nu \bar\nu, \ell^+ \ell^- q \bar q, \nu \bar\nu q \bar q$ 
in the mass range 140 - 1000 GeV \cite{Aad:2015kna}. It is needless to mention that 
better tagging efficiency leads to 4$\ell$ being the most sensitive channel. 
In absence of any significant excess 95\% C.L upper limit on 
$\sigma \times {\rm Br}(H \to ZZ)$ for the two production modes of the 
Higgs boson ($ggF$ and $VBF$) are obtained by combining all the 
four search channels with the narrow width approximation. 
   
The same topology has been searched by the CMS collaboration using 
(5.1 + 19.7 $\ifb$) data collected at $\sqrt s$ = 7 \& 8 TeV LHC run focusing 
in the mass range 145 $< m_H <$ 1000 GeV \cite{Khachatryan:2015cwa}. 
The 95\% C.L exclusion limits for different final states are also presented. Recently 
the ATLAS collaboration has performed a search for heavy Higgs with 
$H \to ZZ \to \ell^+ \ell^- q \bar q$ final state using $3.2~\ifb$ run-II data. Non-observation 
of signal leads to 95\% C.L limits in $\sigma \times {\rm BR}$ in the entire 
mass range 300 - 1000 GeV \cite{atlas_H_ZZ_13tev}. 

\subsubsection{Search for $H$ with $hh$ final states}
\label{sec:Hhh_8}

The heavy Higgs decay to pair of SM Higgses $\rm H\rightarrow hh$ is significant 
only in the low $\tan\beta$ ($\leq 10$) and low to moderate values of 
$M_{A} \leq 400~{\rm GeV}$. At higher values of $M_{A}$ and low values of 
$\tan\beta$, the $H \rightarrow t\bar{t}$ channel dominates, thereby all 
other decay modes get highly suppressed. Depending on the choice of the 
model parameters, production cross section of 
single CP even Higgs can be up to two orders of magnitude larger compared to the 
direct $hh$ production (see Table 1 of Ref.\cite{Bhattacherjee:2014bca}) 
and it can also have non-trivial effects on the self-coupling 
measurement of the 125 GeV Higgs \cite{Bhattacherjee:2014bca}. 
The search for resonant production of a pair of SM Higgs has been done by both the ATLAS and CMS 
collaborations with 8 TeV data assuming various final state topologies like 
$\rm b\bar{b}b\bar{b}$ \cite{Aad:2015uka, Khachatryan:2015yea}, $\rm b\bar{b}\gamma\gamma$ 
\cite{Aad:2014yja, Khachatryan:2016sey}, $\rm b\bar{b}\tau\tau$ \cite{Aad:2015xja} 
and $\rm W W \gamma\gamma$ \cite{Aad:2015xja}. In fact, in Ref.\cite{Aad:2015xja}, 
ATLAS has combined the results for the four above mentioned channels and presented 
the observed and expected 95\% C.L upper limits on $\sigma \times {\rm Br}(H \to hh)$ 
at $\sqrt s$ = 8 TeV (see Fig.6 of Ref.\cite{Aad:2015xja}). Among the four decay modes, 
$\rm hh \to b\bar{b}b\bar{b}$ dominates because of large branching ratio of 
$h \to b \bar b$. However, the reconstruction of the $\rm b\bar{b}b\bar{b}$ final 
state presents a formidable complexity because of the large QCD multi-jet background. 
Thus, less dominant modes like $\rm b\bar{b}\gamma\gamma$ channel become the 
most sensitive in low and relatively heavy (~500 GeV) mass region, as it has 
comparatively less background and better mass resolution. The CMS 
collaboration has investigated both the channels, $H \rm \rightarrow h h \rightarrow b\bar{b}b\bar{b}$ 
and $H \rm \rightarrow h h \rightarrow b\bar{b}\gamma\gamma$, for the mass range 
260 GeV $\rm \leq M_{H} \leq $ 1.1 TeV using LHC-8 data \cite{Khachatryan:2015yea, Khachatryan:2016sey}. 
The ATLAS analyses focused for two different mass regions, for $\rm b\bar{b}b\bar{b}$ they 
restrict themselves in 500 - 1500 GeV, while for $\rm b\bar{b}\gamma\gamma$ they target the 
260 - 500 GeV mass regime \cite{Aad:2015uka,Aad:2014yja}. In absence of signal events, both 
ATLAS and CMS have put 95\% C.L upper limits on the production cross section times branching fraction 
of the heavy resonance for different mass range. 

\subsubsection{Search for $H/{A}$ with $\tau^{+}\tau^{-}$ final states}
\label{sec:Htautau_8}

The search analysis with $H/{A} \to \tau^{+}\tau^{-}$ is one of the 
most important and most promising channels to constrain the MSSM 
parameter space. The LHC run-I data has already severely constrained 
the MSSM parameter space with large to moderate values of $\tan\beta$, say 
$\gtrsim$~15 using $H\to \tau^+ \tau^-$ channel. The reason why regions with 
large values of $\tan\beta$ are excluded is primarily because of the heavy 
Higgs $H,A$ coupling with the SM fermions. For example, the 
coupling of $H,A$ with down (up) type fermions $b, \tau$ (t) increases (decreases) 
with $\tan\beta$ and thus the dominant production modes of $H/A$, mainly ggF and 
associated production with $\rm b\bar{b}$, are primarily controlled by $\tan\beta$.  
Both the ATLAS and CMS collaborations have looked for signatures of $H/A$ produced 
via ggF and $b\bar b\Phi$ ($\Phi$= H,A) processes with Higgs decaying to 
$\rm \tau^{+} \tau^{-}$ \cite{Aad:2014vgg,Khachatryan:2014wca} with LHC-8 data. 
Non-observation of significant excess over the SM backgrounds results into 
model-independent 95\% C.L upper limits on 
$\sigma \times {\rm {BR(\Phi \to \tau^+ \tau^-)}}$ for different values of 
$M_{\Phi}$ with $\Phi$ = H/A at 95 $\rm \%$ C.L. 

Recent results from the ATLAS (CMS) experiment has been reported  
with final states containing tau pair using the data recorded during 
2015 at a centre of mass energy of 13 TeV with $\lum = $ 3.2 (2.3) $\ifb$
\cite{atlas_H_tautau_13tev, CMS:2016pkt}. Both the collaborations have 
seen a very good agreement between the data and the SM backgrounds. As a 
result of this agreement 95\% C.L. upper limits on the production cross section 
times branching fraction has been placed. Now, if one translates these experimental 
upper bounds in terms of the model parameters, say for MSSM in $M_{A} -\tan\beta$ 
plane, and compare the run-II data with the existing run-I data, one can 
clearly observe appreciable improvement in run-II data for regions with 
$M_{A} >$ 700 GeV while in rest of the parameter space run-I and run-II are already 
comparable.

\subsubsection{Search for $A$ with $Zh$ final states}
\label{sec:AZh_8}

The decay of pseudoscalar boson $A$ to $Zh$ is kinematically allowed 
and become appreciable with $M_{A}$ below the $t\bar t$ threshold (i.e., $<$ 350 GeV) 
and very low values of $\tan\beta$ ($<$ 10). Both the ATLAS and CMS collaborations 
have investigated this final state topology with $h$ decaying to 
$\rm b\bar b$ (ATLAS has also analyzed $h \to \rm \tau \tau $ mode \cite{Aad:2015wra}) and 
$Z$ boson into a pair of oppositely charged leptons (electrons or muons) with 8 TeV data 
\cite{Aad:2015wra, Khachatryan:2015lba}. Similar sort of analysis has also been 
looked at recently by the ATLAS collaboration with 3.2 $\ifb$ run-II data \cite{atlas_A_Zh_13tev}. 
From all of these analyses, non observation of any signal of $A$ boson leads to an 
upper limit on the production cross section times branching ratio for a wide range of 
pseudoscalar masses.

\subsubsection{Search for $H^{\pm}$ with $\tau\nu$ and $t\bar b$ final states}
\label{sec:CHtaunu_8}

The observation of a charged Higgs boson signal is one of the smoking gun 
signatures of physics beyond the SM. From the direct searches at 
the Large Electron-Positron (LEP) collider, charged Higgs masses have to be greater 
than around 80 GeV \cite{lepsusy}. At the LHC, the search strategies 
of charged Higgses crucially depend on its mass, in other words, when the charged Higgs is light 
($M_{H^{\pm}} < M_{\rm top}$) then it is primarily produced from the 
$t \bar t$ process and decays into $\tau\nu_{\tau}$ final states. However, for 
heavy $H^{\pm}$ ($M_{H^{\pm}} > M_{\rm top}$), they are mainly produced in 
associated production with top and bottom quarks i.e., $p p \to t b H^{\pm}$. 
 
Both the ATLAS and CMS collaborations have searched for light $H^{\pm}$, 
produced through $t\bar t$ processes and decaying to $c\bar s$ and 
$\tau\nu_{\tau}$ final states with LHC-8 data 
\cite{Aad:2014kga, Khachatryan:2015uua, Khachatryan:2015qxa}. Search for heavy 
charged Higgs via $H^\pm \to t \bar b $ channel with multi $b$-jets in the 
final states has also been studied \cite{Aad:2015typ, Khachatryan:2015qxa}. Again, 
non observation of any excess over the SM predictions leads to 95\% C.L. 
exclusion limits on the production cross-section times branching ratios 
for different values of $M_{H^\pm}$. From the exclusion limits, we find 
that regions with very high $\tan\beta$ ($>$ 20-25) are only severely constrained 
for relatively large values of charged Higgs mass. Recently ATLAS collaboration 
has presented the early run-II results for $H^{\pm}$ decaying into 
$\tau\nu$ final states with $\lum$ = 3.2 $\ifb$ dataset corresponding to 
$\rm \sqrt{s} = 13~{\rm TeV}$ \cite{Aaboud:2016dig}. However, similar to run-I 
data, they have again placed 95\% C.L. upper limits on $\sigma \times {\rm Br}$ 
which are very much similar to the existing 8 TeV bounds.

\subsubsection{Current LHC bounds for electroweakinos}

The relatively smaller production cross section of the electroweak sparticles, namely 
charginos, neutralinos and sleptons, results into comparatively weaker 
bounds on their masses and couplings compared to the colored SUSY particles. 
However with the 8 TeV data, the ATLAS and CMS collaborations 
have already placed strong constraints on the masses/couplings through 
direct searches of the electroweakinos \cite{atlas_ew_3l,atlas_ew_2l,atlas_ew_tau,atlas_ew_higgs, 
atlas_ew_summary,cmsew1,cmsew2}. The mass bounds obtained by the ATLAS and CMS 
are mostly based on simplified SUSY scenarios. However we know that these bounds 
crucially depend on the hierarchy between the slepton and the gaugino masses, 
composition of the electroweakinos and most importantly on 
their branching ratios\footnote{In most of the analysis, 
LHC collaborations assume 100\% branching ratio into one particular
decay mode.}. Majority of these searches assume a bino-like LSP, while 
the second lightest neutralino ($\lsptwo$) and the lighter chargino 
($\chonepm$) to be purely wino-like 
\cite{atlas_ew_3l,atlas_ew_2l,atlas_ew_tau,atlas_ew_higgs, atlas_ew_summary,cmsew1,cmsew2}. 
In fact, the most significant bound is obtained from the $\chonepm \lsptwo$ pair 
production process with trilepton plus missing energy ($\met$) signature \cite{atlas_ew_3l,cmsew2}.

In our 
work we assume sleptons to be sufficiently heavy (masses around 3 TeV), 
and thus the electroweakinos mainly decay via $W,Z,h$ bosons. 
With decoupled scenario the limits become much weaker compared to the 
scenarios with intermediate lighter sleptons\footnote{In a recent study 
\cite{ucmc2} it has been shown that LHC constraints are significantly 
weaker in models with higgsino dominated $\chonepm, \lsptwo$ and $\lspthree$; 
compared to the scenarios studied by the LHC collaborations with 
wino-dominated $\chonepm, \lsptwo$.}. For wino-dominated $\chonepm$ and $\lsptwo$ 
decaying 100\% to $\chonepm\rightarrow\lspone W^{\pm}$ and $\lsptwo\rightarrow\lspone Z$ and 
with decoupled sleptons, the mass limits on charginos obtained by the 
ATLAS and CMS collaborations are 350 GeV for a massless LSP and 305 GeV 
for $\mlspone$ = 125 GeV. However, note that, the limits entirely vanish for $\mlspone >$ 125 GeV. 
The exclusion limits are weakest when $\lsptwo$ decays via the 
{\it spoiler} mode ($h \lspone$) \cite{atlas_ew_higgs, atlas_ew_summary,cmsew2}. 
Moreover, degenerate chargino-neutralino masses up to 148 GeV are 
excluded for LSP masses up to 20 GeV when the decay modes of neutralinos involve a 
Higgs boson through trilepton + $\met$ signature. 
Furthermore, the ATLAS collaboration 
has also presented the electroweakino searches with $1\ell + 2b$ and $1\ell + 2\gamma$ 
($\ell = e,\mu$)final states \cite{atlas_ew_higgs} for simplified models with the 
assumption BR($\lsptwo\rightarrow\lspone h$) = 100\%. The limit obtained from 
$1\ell + 2\gamma$ ($\ell = e,\mu$) channel is little bit stronger than the trilepton bounds. However, it is to be 
remembered that the sharing of the branching ratios between $\lspone h$ and $\lspone Z$ 
can modify the exclusion limits considerably, for details we refer a recent study \cite{Choudhury:2016lku}.

In this work, we consider the present LHC constraints\footnote{We do not consider the 
CMS constraints as we find them comparable with that of ATLAS collaboration.}on 
the electroweakinos via the $3\ell + \met$, $1\ell + 2b +  \met$ and 
$1\ell + 2\gamma + \met$ ($\ell = e,\mu$) channels obtained by the ATLAS collaborations 
with the LHC-8 data. To study\footnote{For details about the implementation and validation of our analysis, 
we refer \cite{Choudhury:2016lku}.} the impact of these constraints on our parameter space, we first 
calculate the signal and background efficiencies for some representative benchmark points 
(see next section), and then using the NLO production cross-sections as obtained from the {\tt PROSPINO} 
\cite{Beenakker:1996ed} we estimate the effective cross sections and then make sure that our numbers 
lie below the 95\% C.L. upper limits already presented by the ATLAS collaboration in the 
three above-mentioned channels.


\section{Search strategies and Future limits}
\label{sec3:collider}

From the previous section, we find that the 
branching fraction of the heavy Higgses to non-SM 
`ino' pairs can be significantly large. Moreover, 
the charginos and neutralinos produced from the decay 
of heavy Higgses can themselves also undergo long 
cascade decays depending upon the choice of MSSM parameters. 
As such, a myriad of possible cascade decay modes are possible 
corresponding to various final state topologies. For example, 
one can have a very simple cascade decay of the form 
$H \to \lsptwo \lspone, \, \lsptwo \to Z/h \lspone$, resulting 
in a mono-Z/h + $\met$ final state or, 
$H^{\pm} \to \chonepm \lspone, \, \chonepm \to W^{\pm} \lspone $, 
resulting in mono-W + $\met$ final state. However, one can also 
encounter relatively complex cascade decays with multiple decay chains 
in between, for example, 
$ H \to \lsptwo \lspthree, \, \lsptwo \to Z \lspone , \, \lspthree \to W^{\pm} \chonepm $,  
which may eventually lead to final states with multiple leptons and jets along 
with large missing energy. As we have already mentioned, one 
can have a significant branching fraction for these cascade decays 
by making judicious choices of the parameters. Proper understanding 
of these cascade decays with various possible final state 
topologies can be considered as a direct probe of the MSSM 
Higgs sector. In this work, we perform a detailed collider analysis 
of the MSSM heavy Higgses focusing the mono-X + $\met$ ($X = W,Z$) and 
trilepton + $\met$ channels in the context of 14 TeV 
high luminosity run of LHC. Here we assume that the SM gauge bosons 
$W/Z$ decay leptonically i.e., $W \to \ell \nu$ and $ Z \to \ell \ell $ with 
$\ell = e,\mu$, in order to reduce the SM backgrounds significantly. 
The hadronic decay modes might also be important, however, 
we do not consider them in this work. Searches through mono-Higgs plus $\met$ 
also exist literature, for details we refer \cite{swagato}.

\subsection{Benchmark points}
\label{bp}

\begin{table}[htb!]
\begin{center}
\begin{tabular}{|c|c||c||c|c|} \hline
 Benchmark & Parameters (GeV) & Mass (GeV) & Processes & Branching \\ 
 Points& & & & Fraction\\ \hline\hline
 	& $ M_{A} = 591.2, \quad M_{1} = 127.1,$ &  $ M_{\chi_{1}^{0}}\,=\,119.7 $  & $ H \rightarrow \chi_{2}^{0}\,\chi_{1}^{0}$ & $  4.58\% $ \\ 
BP-1 	& $ M _{2} = 900, \quad \mu = 237.2, $ & $ M_{\chi_{2}^{0}}\,=\,241.8 $ & $ H \rightarrow \chi_{3}^{0}\,\chi_{1}^{0}$ & $  10.14\% $ \\
	 & $ \tan{\beta} = 15~ ,\quad A_{t} = 1890, $ & $ M_{\chi_{3}^{0}}\,=\,241.8 $ & $ A \rightarrow \chi_{2}^{0}\,\chi_{1}^{0}$ & $  9.23\% $ \\
	 & $ m_{\tilde{Q}_{3l}} = 4160, \quad m_{\tilde{t}_{R}} = 6520, $ & $ M_{\chi_{4}^{0}}\,=\,907.4 $ & $ A \rightarrow \chi_{3}^{0}\,\chi_{1}^{0}$ &  $ 4.65\% $  \\
 	& $ m_{\tilde{b}_{R}} = 2280, \quad A_{b} = A_{\tau} = 0 $ & $ M_{\chi_{1}^{\pm}}\,=\,234.3 $ & $ \chi_{2}^{0} \rightarrow Z\, \chi_{1}^{0}$ & $ 100\%$ \\
	 & $ M_{3} = 2960$ &  $ M_{\chi_{2}^{\pm}}\,=\,907.4 $ & $ \chi_{3}^{0} \rightarrow Z\, \chi_{1}^{0}$ & $ 100\%$ \\ 
 & & $ M_{H} = 591.3 $ & & \\ 
 & & $ M_{H^{\pm}} = 596.8 $ & & \\ \hline
  & $ M_{A} = 550, \quad M_{1} = 80,$ &  $ M_{\chi_{1}^{0}}\,=\,77.2 $  & $ H \rightarrow \chi_{2}^{0}\,\chi_{1}^{0}$ & $ 4.82\% $ \\ 
BP-2 & $ M _{2} = 900, \quad \mu = 350, $ & $ M_{\chi_{2}^{0}}\,=\,347.8 $ & $ H \rightarrow \chi_{3}^{0}\,\chi_{1}^{0}$ & $  13.93\% $ \\
 & $ \tan{\beta} = 8.5~ ,\quad A_{t} = 3770, $ & $ M_{\chi_{3}^{0}}\,=\,353.6 $ & $ A \rightarrow \chi_{2}^{0}\,\chi_{1}^{0}$ & $  14 .14\% $ \\
 & $ m_{\tilde{Q}_{3l}} = 3380, \quad m_{\tilde{t}_{R}} = 9040, $ & $ M_{\chi_{4}^{0}}\,=\,908.5 $ & $ A \rightarrow \chi_{3}^{0}\,\chi_{1}^{0}$ &  $ 3.89\% $  \\
 & $ m_{\tilde{b}_{R}} = 2820, \quad A_{b} = A_{\tau} = 0 $ & $ M_{\chi_{1}^{\pm}}\,=\,345.1 $ & $ \chi_{2}^{0} \rightarrow Z\, \chi_{1}^{0}$ &$ 24.25\%$ \\
 & $ M_{3} = 8900$ &  $ M_{\chi_{2}^{\pm}}\,=\,908.5 $ & $ \chi_{3}^{0} \rightarrow Z\, \chi_{1}^{0}$ & $ 83.56\%$ \\ 
 & & $ M_{H} = 550.6 $ & & \\ 
 & & $ M_{H^{\pm}} = 556.0 $ & & \\ \hline
 & $ M_{A} = 600,\quad M_{1} = 950,$ &  $ M_{\chi_{1}^{0}}\,=\,158.2 $  & $ H \rightarrow \chi_{1}^{\pm}\,\chi_{2}^{\mp} $ & $ 23.39\% $ \\ 
BP-3 & $ M _{2} = 178.2, \quad \mu = 286.1, $ & $ M_{\chi_{2}^{0}}\,=\,292.7 $ & $ A \rightarrow \chi_{1}^{\pm}\,\chi_{2}^{\mp} $ & $ 16.70\% $ \\
 & $ \tan{\beta} = 21~ ,\quad A_{t} = 4320, $ & $ M_{\chi_{3}^{0}}\,=\,310.3 $ & $ \chi_{2}^{\pm} \rightarrow W^{\pm} \,\chi_{1}^{0}$ & $  43.48\% $ \\
 & $ m_{\tilde{Q}_{3l}} = 3370, \quad m_{\tilde{t}_{R}} = 4230, $ & $ M_{\chi_{4}^{0}}\,=\,952.3 $ & $ H \rightarrow \chi_{1}^{0}\,\chi_{2}^{0}$ &  $ 8.30\% $  \\
 & $ m_{\tilde{b}_{R}} = 5330, \quad A_{b} = A_{\tau} = 0 $ & $ M_{\chi_{1}^{\pm}}\,=\,159.0 $ & $ H \rightarrow \chi_{1}^{0}\, \chi_{3}^{0}$ & $ 1.30 \% $ \\
 &$ M_{3} = 7100$ &  $ M_{\chi_{2}^{\pm}}\,=\,316.8 $ & $ A \rightarrow \chi_{1}^{0} \chi_{2}^{0}$ & $  3.05\% $  \\ 
 & &$ M_{H} = 600.0$ & $A \rightarrow \chi_{1}^{0} \chi_{3}^{0}$ & $ 4.10\% $ \\
 & & $ M_{H^{\pm}} = 605.5 $ &  $\chi_{2}^{0} \rightarrow \chi_{1}^{\pm}\,W^{\mp}$ & $ 73.32\%$ \\
 & & & $ \chi_{3}^{0} \rightarrow \chi_{1}^{\pm}\,W^{\mp}$ & $81.06\%$ \\ \hline
 & $ M_{A} = 657.5, \quad M_{1} = 159.5,$ &  $ M_{\chi_{1}^{0}}\,=\,145.5 $  & $ H \rightarrow \chi_{1}^{\pm}\,\chi_{2}^{\mp} $ & $  28.971\% $ \\ 
BP-4 & $ M _{2} = 337.2, \quad \mu = 236.6, $ & $ M_{\chi_{2}^{0}}\,=\,230.7 $ & $ A \rightarrow \chi_{1}^{\pm}\,\chi_{2}^{\mp} $ & $ 15.8\% $ \\
 & $ \tan{\beta} = 23~ ,\quad A_{t} = 1290, $ & $ M_{\chi_{3}^{0}}\,=\,248.9 $ & $ \chi_{2}^{\pm} \rightarrow W^{\pm} \,\chi_{3}^{0}$ & $ 21.85\% $ \\
 & $ m_{\tilde{Q}_{3l}} = 9590, \quad m_{\tilde{t}_{R}} = 1920, $ & $ M_{\chi_{4}^{0}}\,=\,387.6 $ & $ \chi^{0}_{3} \rightarrow Z \chi_{1}^{0}$ &  $ 100\% $  \\
 & $ m_{\tilde{b}_{R}} = 2600, \quad A_{b} = A_{\tau} = 0 $ & $ M_{\chi_{1}^{\pm}}\,=\,221.4 $ & $ H \rightarrow \chi_{3}^{0}\, \chi_{4}^{0}$ & $ 8.89\%$ \\
 &$ M_{3} = 6180$ &  $ M_{\chi_{2}^{\pm}}\,=\,387.3 $ & $ A \rightarrow \chi_{3}^{0} \chi_{4}^{0}$ & $ 0.39\%$ \\ 
 & &$ M_{H} = 657.5$ & $ \chi_{4}^{0} \rightarrow W^{\pm} \chi_{1}^{\pm}$ & $ 67.78\%$ \\
 & & $ M_{H^{\pm}} = 662.4 $ &   & \\ \hline
\end{tabular}
\end{center}

\caption{Input parameters, masses of heavy Higgses and electroweakinos, and 
branching fraction of relevant processes for benchmark points. Here 
all the input mass parameters and output masses are in GeV.}
\label{tab:mono_bpt}
\end{table}

To present the collider analysis we choose four optimized  
benchmark points (BP-1 to BP-4) suitable for the above-mentioned 
three search strategies. All the model parameters, masses of heavy 
Higgses and electroweakinos and the relevant branching ratios are 
summarized in Table \ref{tab:mono_bpt}. For all the four 
benchmark points, heavy Higgs masses lie in the mass range of 500 - 700 GeV. 
The first two benchmark points, BP-1 and BP-2, are optimized 
for the mono-Z + $\met$ searches, where the lightest chargino and 
the second/third lightest neutralinos are higgsino dominated 
and LSP is bino dominated. In BP-1, BR$(\lsptwo,\lspthree \to Z \lspone)$ = 100\% 
and the total contribution coming from H/A to $\lsptwo \lspone, \lspthree \lspone$ 
are about 30\%, while in BP-2, the decay modes $\lsptwo,\lspthree \to h \lspone$ 
open up and consequently the branching ratios to $Z \lspone$ get reduced. Our third 
benchmark point, BP-3, is optimized for the mono-W + $\met$ searches where 
LSP is mixture of substantial wino and higgsino components while 
$\lsptwo, \lspthree$ are higgsino dominated. Here $\chonepm$ is 
primarily wino but has significant higgsino component. In Table \ref{tab:mono_bpt}, all 
the relevant branching ratios of $ H/A \rightarrow \chi_{1}^{\pm}\,\chi_{2}^{\mp}, \lspone \chi_{2,3}^{0}$ 
are summarized. In BP-4, the input parameters $M_1, M_2$ and $\mu$ all are 
relatively light and close to each others, and hence the charginos and neutralinos are 
mixed states of bino, wino and higgsino states. Several possible long cascade 
decays of these electroweakinos (see last column of Table \ref{tab:mono_bpt}) leads 
to final state topologies with trileptons plus missing energy. 
 

\begin{table}[!htb]
\begin{center}
\begin{tabular}{|c||c|c|c|c||c|c|c|c|}
\hline
 & \multicolumn{8}{c|}{Production Cross-section in fb } \\ \cline{2-9}
 & \multicolumn{4}{c||}{ 8 TeV} & \multicolumn{4}{c|}{ 13(14) TeV} \\ \cline{2-9}
 & BP-1 & BP-2 & BP-3 & BP-4 & BP-1 & BP-2 & BP-3 & BP-4 \\ 
\hline\hline
 H & 65  & 41  & 114  & 21 & 290(356)  & 171(207)  & 517(635)  & 100(129)\\
 A & 70  & 51  & 119  & 24 & 311(380)  & 208(252)  & 534(657)  & 114(140)\\ 
 $\rm H^{\pm}$ & 1.1  & 0.8  & 1.9  & 0.4 & 7.3(9.4) & 5.12(6.5)  & 12.9(16.8)  &  3.2(4.2) \\ \hline
\end{tabular}

\caption{Production cross-sections of the heavy Higgses $H,A$ and $\rm H^{\pm}$ 
for $\rm \sqrt{s} = 8, 13 ~{\rm and}~14 ~{\rm TeV}$.  Higgses 
are predominantly produced through ggF process. Charged Higgs production cross-sections are 
small, so we ignore them in our analysis.}
\label{tab:mono_prod_cs}
\end{center}
\end{table}

In Table \ref{tab:mono_prod_cs} we present the production cross-sections of 
the MSSM heavy Higgses $H,A$ and $\rm H^{\pm}$ for center-of-mass energies 
$\rm \sqrt{s} = 8, 13 ~{\rm and}~14 ~{\rm TeV}$. We assume that these Higgses 
are predominantly produced via the gluon-gluon fusion process. From the last row of 
Table~\ref{tab:mono_prod_cs}, it is clear that the charged Higgs production 
cross-sections are much smaller compared to the $H,A$ cross-sections, and 
hence we ignore them in rest of our analysis. Once we know the cross-sections 
of these Higgses and also branching ratios to different decay modes, we 
can calculate the quantity $\sigma \times {\rm BR}$ for all the benchmark 
points and then compare our results with the present experimental bounds. In 
Table \ref{tab:Higgs_UL}, we display the observed 95\% C.L. upper limits 
on the production cross-section times branching ratios for the heavy Higgses 
obtained by the ATLAS and CMS collaborations using the run-I and run-II data. We 
also quote the predictions for a given decay mode associated to all the four 
benchmark points. Here, we consider the following Higgs decay modes: 
$H/A \to  \gamma\gamma, WW, ZZ, hh, \tau\tau$;  $A \to Zh$; 
$H^{\pm} \to \tau\nu, t\bar b$. The quantity $\sigma \times {\rm Br}(H/A \to X X)_{ATLAS/CMS}^{UL}$ 
denotes the 95\% C.L. upper limit on the production cross-section times 
branching ratios observed by the ATLAS/CMS collaborations, and 
$\sigma \times {\rm Br}(H/A \to X X)$ is the same but for our representative 
benchmark points for a generic heavy Higgs decay $H/A \to X X$ channel. We observe that, in 
most of the cases, the limits from LHC run-I data is more stringent than 
the early run-II data and, except for the $H/A \to \tau \tau$ channel, the 
observed upper limits are few order of magnitude smaller than the predicted values 
for the benchmark points.  
 
\begin{table}[htb!]
\small{
\begin{tabular}{|c|c|c|c|c|c|c|c|c|}
\hline
Modes &\multicolumn{4}{c|}{$\rm \sqrt{s} = 8~TeV (fb)$} &\multicolumn{4}{c|}{$\rm \sqrt{s} = 13~TeV (fb)$  } \\  \cline{2-5} \cline{6-9}
 								& BP-1	& BP-2	& BP-3	& BP-4	& BP-1	& BP-2	& BP-3	& BP-4	\\
\hline
\hline
 \footnotesize{$\sigma \times {\rm Br}(H/A \to \gamma \gamma)_{ATLAS}^{UL}$}\cite{Aad:2014ioa,Aaboud:2016tru}	&  0.9	&  1.8	&  0.6	&  -	&  3.0 	&  3.6	&  3.3	&  2.0	\\
 \footnotesize{$\sigma \times {\rm Br}(H/A \to \gamma \gamma)_{CMS}^{UL}$}\cite{Khachatryan:2015qba,Khachatryan:2016hje}	&  1.4	&  1.2	&  1.0	&  0.5	&  5.1	&  4.6	&  3.7	&  1.3	\\
 \footnotesize{$\sigma \times {\rm Br}(H/A \to \gamma \gamma)$} 	& $\sim 10^{-6}$ & $\sim 10^{-4}$  &  $\sim 10^{-4}$ 	& $\sim10^{-6}$  &  $\sim 10^{-4}$  	& $\sim 10^{-3}$ 	& $\sim 10^{-3}$  	&  $\sim 10^{-4}$ \\

\hline
\hline
 \footnotesize{$\sigma \times {\rm Br}(H \to WW)_{ATLAS}^{UL}$}\cite{Aad:2015agg,atlas_H_WW_13tev}		&  242.8&  308.6&  229.3&  186.5 & 570.7 	& 817.7 & 543.2	& 457.8  	\\
 \footnotesize{$\sigma \times {\rm Br}(H \to WW)_{CMS}^{UL}$}	\cite{Khachatryan:2015cwa}		&  124.99 	& 148.02  	& 121.28  	& 109.65  	&  - 	& -  	& -  	& -  	\\
 \footnotesize{$\sigma \times {\rm Br}(H \to WW)$} 				& 0.04  	& 0.18  	& 0.01  	& 0.01  	& 0.18  	& 0.76  	& 0.06  	& 0.06  	\\
\hline
\hline
 \footnotesize{$\sigma_{ggH} \times {\rm Br}(H \to ZZ)_{ATLAS}^{UL}$}\cite{Aad:2015kna,atlas_H_ZZ_13tev}		  &23.7 & 30.8  & 21.2  & 26.2  &  329.8 & 456.4  & 304.1  &  195.8 	\\
 \footnotesize{$\sigma_{ggH} \times {\rm Br}(H \to ZZ)$} 	&  $\sim10^{-3}$ 	& 0.02  	&  $\sim 10^{-4}$ 	& $\sim10^{-5}$ &  $\sim10^{-3}$ 	&  0.11  	&   $\sim 10^{-3}$	 & $\sim 10^{-3}$	\\ \hline
 \footnotesize{$\sigma \times {\rm Br}(H \to ZZ)_{CMS}^{UL}$}\cite{Khachatryan:2015cwa}	 &  128.96 	&  140.01 	& 125.32  	& 117.71  	&  - 	& -  	& -  	& -  	\\
 \footnotesize{$\sigma \times {\rm Br}(H \to ZZ)$} 	&  0.02 	& 0.09  	& $\sim10^{-2}$  	& $\sim10^{-3}$ &  0.09 	&  0.37	& 	0.03  & 0.03	\\ 
\hline
\hline
 \footnotesize{$\sigma_{ggF} \times {\rm Br}(H \to hh)_{ATLAS}^{UL}$}	\cite{Aad:2015xja}	  & 87.8 & 121.1  & 79.1  	& 42.3  	&-   	& -  	& -  	& -  	\\
 \footnotesize{$\sigma_{ggF} \times {\rm Br}(H \to hh)$} 		 &  0.01 	& 0.2  	& 0.002  	& 0.006  	& 0.04  	& 0.78  	& 0.009  	& 0.024   	\\
 \hline
 \footnotesize{$\sigma \times {\rm Br}(H \to hh \to \gamma \gamma b\bar{b})_{CMS}^{UL}$}\cite{Khachatryan:2016sey}			  &  1.19 	& 1.47  	& 1.13  	& 0.77  	&-   	&-   	& -  	&  - 	\\
 \footnotesize{$\sigma \times {\rm Br}(H \to hh \to \gamma \gamma b\bar{b})$} 		 &  $\sim 10^{-4}$ 	& $\sim 10^{-4}$   	& $\sim 10^{-4}$  	& $\sim 10^{-4}$   	&  $\sim 10^{-3}$  	&  $ \sim 10^{-3}$ 	&  $\sim 10^{-3}$  	&  $\sim 10^{-4}$ 	\\
\hline
\hline
 \footnotesize{$\sigma_{gg\phi} \times {\rm Br}(\phi \to \tau \tau)_{ATLAS}^{UL}$}\cite{Aad:2014vgg,atlas_H_tautau_13tev} 	  &  19.1 	& 21.9  	& 19.0  	& 15.6  	& 112.0  &  148.3 	& 104.1  & 83.8   	\\
 \footnotesize{$\sigma_{gg\phi} \times {\rm Br}(\phi \to \tau \tau)_{CMS}^{UL}$}\cite{Khachatryan:2014wca, CMS:2016pkt}	  &  23.1 	& 29.6  	& 22.4  	& 17.5  	& 85.7  & 96.5 & 84.6  & 75.2   	\\
 \footnotesize{$\sigma_{ggH+ggA} \times {\rm Br}(H/A \to \tau \tau)$} 		  & 1.46  	& 2.24  	& 1.14  	& 0.31  	& 5.62  	& 8.38  	&  4.44 & 1.29  	\\
\hline
 \footnotesize{$\sigma_{bb\phi} \times {\rm Br}(\phi \to \tau \tau)_{ATLAS}^{UL}$}\cite{Aad:2014vgg,atlas_H_tautau_13tev}	  &  19.1  	& 21.9  	& 18.6  	& 14.9  	& 106.7 	& 147.8 	& 97.6  	& 80.1  	\\
 \footnotesize{$\sigma_{bb\phi} \times {\rm Br}(\phi \to \tau \tau)_{CMS}^{UL}$}\cite{Khachatryan:2014wca, CMS:2016pkt}		  &  22.7 	& 25.7  	& 21.4   & 20.2 	& 72.8  	& 79.0 	& 71.1  	& 93.4   	\\
 \footnotesize{$\sigma_{bbH+bbA} \times {\rm Br}(H/A \to \tau \tau)$} 			  &  11.62 	& 3.51  	& 14.97	& 1.30  	& 52.46  & 15.15  	& 68.28  	& 6.35  	\\
\hline
\hline
 \footnotesize{$\sigma \times {\rm Br}(A \to Zh)_{ATLAS}^{UL}$}\cite{Aad:2015wra, atlas_A_Zh_13tev}		  &  41.4 & 43.1  & 41.0  & 26.3 & 412.6 & 709.9  & 365.3 & 250.3   	\\
 \footnotesize{$\sigma \times {\rm Br}(A \to Zh)_{CMS}^{UL}$}\cite{ Khachatryan:2015lba}	& 83.29 & 131.9 & 67.0  	& - & - 	& - 	& - 	&  -  	\\
 \footnotesize{$\sigma \times {\rm Br}(A \to Zh)$} 				  &   0.04	& 0.2  	& 0.013  &   0.0062	& 0.16  	& 0.8  	& 0.06  	& 0.03  	\\
\hline
\hline
 \footnotesize{$\sigma \times {\rm Br}(H^{\pm} \to \tau\nu)_{ATLAS}^{UL}$}\cite{Aad:2014kga,Aaboud:2016dig}			  &  11.12 & 14.48  	& 9.94  	&  8.84 	& 53.4  	& 66.9  	& 50.3  	& 39.2  	\\
 \footnotesize{$\sigma \times {\rm Br}(H^{\pm} \to \tau\nu)_{CMS}^{UL}$}\cite{Khachatryan:2015qxa}			  &  26.28 	&  30.31 	&  25.97 	& -  	& -  	& -  	& -  	&-   	\\
 \footnotesize{$\sigma \times {\rm Br}(H^{\pm} \to \tau\nu)$} 				  &  0.12 	&  0.05 	& 0.14  	&  0.02 	& 0.76  	& 0.33  	& 0.96  	&  0.12 	\\
\hline
\hline
 \footnotesize{$\sigma \times {\rm Br}(H^\pm \to t\bar b)_{ATLAS}^{UL}$}\cite{Aad:2015typ}	  &  208.7 	& 487.7  	& 238.8  	&  - & -  	& -  	& -  	& -  	\\
 \footnotesize{$\sigma \times {\rm Br}(H^\pm \to t\bar b)_{CMS}^{UL}$}\cite{Khachatryan:2015qxa}		 &   137.1	& 166.1  &  132.7 	& - 	& -  	&-   	& -  	& -  	\\
 \footnotesize{$\sigma \times {\rm Br}(H^\pm \to t\bar b)$} 				  &  0.78 	& 0.59  	& 0.89  	& 0.13  	& 5.09  	& 3.69  	& 5.90  	& 0.99  	\\
\hline
\end{tabular}
  \caption{\small \it  Upper limits on the production cross-section times the branching ratios at 95\% C.L. 
for various decay modes obtained by ATLAS and CMS collaborations and the corresponding values 
for the benchmark points introduced in Table~\ref{tab:mono_bpt}.}
  \label{tab:Higgs_UL}
}
\end{table}

In addition to the heavy Higgs direct search bounds, we also consider 
the limits obtained from the direct searches of the 
charginos and neutralinos at the LHC. Both the ATLAS and CMS 
collaboration have searched for these electroweakinos, and placed 
upper limits on the production cross-sections. We find that the limits 
of ATLAS and CMS collaborations are comparable, and so in our analysis 
we use the ATLAS limits only. All the benchmark points, BP-1 to BP-4, 
posses significant gaugino-higgsino mixing which leads to sizeable 
modifications in electroweakino pair production cross-sections. We calculate 
all the dominant chargino-neutralino pair production cross-sections 
(see Table \ref{tab:EW_limit}) and then following the ATLAS 
analysis \cite{atlas_ew_3l,atlas_ew_higgs}, we estimate the 
cut efficiencies for the three final state topologies 
$3\ell + \met$, $1\ell + 2b + \met$ and $1\ell + 2\gamma + \met$ ($\ell = e,\mu$). 
In Table~\ref{tab:EW_limit}, we present the observed 95\% C.L. upper limits 
on the number of BSM signal events, $N_{BSM}^{Obs.~UL}$, 
by the ATLAS collaborations for the most sensitive signal regions for the 
above-mentioned three channels. In Table~\ref{tab:EW_limit} we also present the 
production cross-sections and $N_{BSM}$ for our benchmark points for 8 TeV LHC  
with $\lum = $ 20.3 $\ifb$, where the quantity $N_{BSM}$ = production
cross-section  $\times$ efficiency $\times$ acceptance $\times$ luminosity 
for a given channel. From the table, it is evident that the models predictions 
are much smaller compared to the observed 95\% C.L. upper limits, and thus all our 
benchmark points are consistent with the updated bounds associated to direct searches 
of the electroweakinos at the LHC.     
 

\begin{table}[!htb]
\begin{center}
\small
\begin{tabular}{|c||c|c|c|c||}
\hline
				 	&BP-1	&BP-2	&BP-3	&BP-4   \\
\hline
\hline
$\sigma_(pp \ra \chonepm \lsptwo)$ 	&91.01  &18.76   &17.75  &131.24  \\
\hline
$\sigma_(pp \ra \chonepm \lspthree)$	&95.57 &17.77  &13.12  &85.55   \\
\hline
$\sigma_(pp \ra \chonepm \lspfour)$ 	&-	 &-	   &-     & 3.32   \\
\hline
$\sigma_(pp \ra \chtwopm \lsptwo)$  	&-	 &-	   &29.35     &3.61    \\
\hline
$\sigma_(pp \ra \chtwopm \lspthree)$ 	&-	 &-	   &39.39     &3.93    \\
\hline
$\sigma_(pp \ra \chtwopm \lspfour)$  	&-	 &-	   &-     & 35.07   \\
\hline
$\sigma_{Total} $   			&187.6  &36.53   &99.61  &262.68   \\
\hline
\hline
$N_{BSM}$ for SR0$\tau$a-bin16 		&0.93	 &0.99	   &0.48    &0.65    \\
($N_{BSM}^{Obs.~UL}$ = 5.2\cite{atlas_ew_3l})&	 &	   &     &    \\
\hline
\hline
$N_{BSM}$ for SR$lbb$-2	 		&0	 &0.76	   &0.1     & 0.05   \\
($N_{BSM}^{Obs.~UL}$ = 5.5\cite{atlas_ew_higgs})&	 &	   &     &    \\
\hline
\hline
$N_{BSM}$ for SR$l\gamma \gamma$-1 	&0	 &0.08	   &0.05     & 0.02   \\
($N_{BSM}^{Obs.~UL}$ = 3.6\cite{atlas_ew_higgs})&	 &	   &     &    \\
\hline
       \end{tabular}
       \end{center}
           \caption{Here the cross-sections (NLO) are in fb. $N_{BSM}^{Obs.~UL}$ stands for 
Observed upper limits on  $N_{BSM}$ at 95 $\%$ CL where $N_{BSM}$ = production
cross-section  $\times$ efficiency $\times$ acceptance $\times$ luminosity. }
\label{tab:EW_limit}
          \end{table}


\subsection{Mono-X plus missing energy}
\label{monoX}

Events with a single $W/Z$ boson plus missing transverse energy 
($\met$) constitute a very clean and distinctive signature in new physics 
searches at the LHC. This topology has been thoroughly analyzed by both 
the ATLAS and CMS collaborations, mainly in the context of DM searches. 
In this paper, we consider two such mono-X + $\met$ channels, namely mono-$Z$ and 
mono-$W$, with both $W,Z$ decaying leptonically ($Z \to ll,\, W\to l \nu, l = e,\mu$ ) 
to search for the MSSM heavy Higgses at the LHC.

Below, we discuss the details of the collider analysis for 
our optimized benchmark points for the above-mentioned mono-X search channels.  
We use {\tt MadGraph (version 2.3.3)} \cite{Alwall:2014hca} to generate the background events 
and {\tt PYTHIA (version 6.428)} \cite{pythia6} for showering and hadronization. 
The production cross-sections of the heavy Higgses $H,A$ have been 
calculated using {\tt SusHi (version 1.5.0)} \cite{sushi}, while for the 
charged Higgs boson $H^{\pm}$ we use {\tt PYTHIA}. To obtain the particle 
spectrum we use {\tt SUSPECT} while {\tt SUSYHIT} 
has been used to calculate the Higgs and SUSY decay widths and branching ratios. 
The signal events have been generated through gluon-gluon fusion process 
using {\tt PYTHIA}. Both the signal and background 
events have been passed through a fast detector simulation using 
{\tt Delphes3 (version-3.3.2)} \cite{deFavereau:2013fsa} using the default ATLAS card. 

\subsubsection{Mono-Z (leptonic) + $\met$ channel}
\label{sec:monoz}

We perform a search for the MSSM heavy Higgses in events with 
a leptonically decaying $Z$ boson ($Z \to \ell^+\ell^-, \ell = e,\mu$) 
produced through cascade decays of the charginos and neutralinos 
in the context of $\rm \sqrt{s} = 14~TeV$ run of LHC with an 
integrated luminosity of 3000~$\ifb$. These events also contain 
significant missing transverse energy coming from the lightest 
neutralinos. The search strategy reported in \cite{Aad:2014vka}, which focused 
on the DM searches, has been followed with suitable modifications 
aimed to optimize the signal significance. This analysis has 
been performed on the first two representative benchmark points, BP-1 and BP-2 
(see Table \ref{tab:mono_bpt}). The relevant decay chains giving rise to 
the above-mentioned final state signature are: $ p\, p \rightarrow H/A,\quad H/A 
\rightarrow \wt\chi_{2,3}^0 \lspone, \quad  \wt\chi_{2,3}^0 \rightarrow \lspone Z $.

The events are selected with two same flavour opposite sign (SFOS) 
isolated leptons (electrons or muons) with $p_T$ greater than 20 GeV. 
Candidate electrons(muons) are required to be within pseudo-rapidity 
range $\rm |\eta| < 2.47 (2.5)$. For an electron or muon to be isolated, 
the scalar $p_T$ sum of all stable particles with $p_{T} > 1$~GeV 
present within a cone of radius $\Delta R = 0.2$ around the direction of 
candidate lepton should be less than $10\%$ of $p_T$ of the candidate 
lepton, where $\Delta{R} = \sqrt{{(\Delta \phi)}^{2} + {(\Delta \eta)}^{2}}$ 
with $\phi$ being the azimuthal angle. The azimuthal angle between 
the dilepton system and $\met$ direction, $\Delta{\phi}_{p_{T}^{\ell\ell},\met}$, is 
required to be greater than 2.5, where $p_{T}^{\ell\ell}$ is the momentum of 
the dilepton system. In addition, the absolute value of pseudo-rapidity 
of the dilepton system, $|\eta_{\ell\ell}|$, must lie within 2.5 and the 
invariant mass of the SFOS pair is required to be within 
$M_{Z} \pm 15~{\rm GeV}$, where $ M_{Z}$ is the Z boson mass. Events 
with one or more jets with $p_{T} $ greater than 25 GeV are rejected. To further 
reduce the SM backgrounds, we define a kinematic variable $\xi$ constructed 
using the dilepton $p_T$ and $\met$ as $| p_{T}^{\ell\ell} - \met |/p_{T}^{\ell\ell}$ 
where $\ell = e,\mu$.  
\begin{table}[!htb]
\begin{center}
\begin{tabular}{|c|c|}
\hline
 Signal Regions & Selection Cuts \\ 
\hline
SRA1 & ${\PMET > 125~{\rm GeV}} \quad \& \quad \xi < 0.3$ \\
SRB1 & ${\PMET > 150~{\rm GeV}} \quad \& \quad \xi < 0.5$ \\
\hline
\end{tabular}
\caption{Signal regions SRA1 and SRB1 defined using the missing energy ($\met$) and 
transverse momentum of the dilepton system where we define 
$\xi = \frac{|p_{T}^{\ell\ell} - \PMET|}{p_{T}^{\ell\ell}}$. }
\label{tab:SRs1}
\end{center}
\end{table}

The dominant SM background, an irreducible background too, is 
$ p p \to ZZ \rightarrow \ell^{+} \ell^{-}\,\nu\,\bar{\nu} ( \ell = e,\mu)$. However, 
processes like $ p p \to W^{+}W^{-} \rightarrow \ell^{+}\,\nu \,\ell^{-}\,\bar{\nu}$, 
$ WZ \rightarrow \ell \,\nu \,\ell^{+}\,\ell^{-}$ and 
$ ZZ \rightarrow \ell^{+}\,\ell^{-}\,\ell^{+}\,\ell^{-}$ also contribute to 
the SM backgrounds when additional leptons get misidentified or remain 
unreconstructed. Two signal regions, SRA1 and SRB1, are constructed using 
specific choices of cuts on $\met$ and $\xi$, see Table \ref{tab:SRs1}. 
The signal region SRB1 is motivated from the ATLAS 
analysis \cite{Aad:2014vka} where $\met$ is large. The signal region SRA1 is based 
on our re-optimization of the signal significances with relatively smaller 
values of $\met$ and $\xi$. The number of signal and background events 
corresponding to each signal regions are shown in Table \ref{tab:monoZ_signi}. 
For both SRA1 and SRB1, we find that one can obtain $5\sigma$ 
statistical significance at the 14 TeV LHC with 3000~$\ifb$ of luminosity, 
here statistical significance has been calculated as $S/\sqrt{B}$ where S(B) 
is the number of signal (background) events.

Here we would like to mention that in order to perform this search, 
the lower limits on the $\met$ selection cuts need to be restricted 
to relatively lower values $( < 200~{\rm GeV})$ as the signal yield 
becomes statistically insignificant in the higher $\met$ regime. 
The results of this analysis indicate towards a possibility 
to marginally discover/exclude the heavy Higgses at high luminosity LHC 
run. However, we would like to mention here that the signal significances 
have been calculated assuming zero systematic uncertainty. As the systematic 
uncertainty comes into play, the signal significance will get 
significantly reduced and will go down much below the $5\sigma$ discovery limit.

\begin{table}[!htb]\centering
\begin{tabular}{|c|cc|c c|cc|}
\hline
Signal & \multicolumn{2}{c|}{Signal} & \multicolumn{2}{c|}{Backgrounds} & \multicolumn{2}{c|}{Significance}\\
\cline{2-7}
Regions   & BP-1 & BP-2   & $ Z Z $  & $ W Z $ &  BP-1 & BP-2 \\
\cline{1-7}
      SRA1   & 921  & 804 &  15077   &  5738  & 7.45  & 6.50 \\
      SRB1   & 506  & 619 &  9187  &  3152   & 5.24  & 6.41 \\
\hline
\end{tabular}
\caption{\small \it Number of signal and background events at the 14 TeV run 
of LHC with 3000~$\ifb$ of luminosity obtained after the imposition of our selection cuts 
for the mono-$Z+\PMET$ channel. Here we focus on the first two representative 
benchmark points BP-1 and BP-2. }
\label{tab:monoZ_signi}
\end{table}


%

\subsubsection{Mono-W (leptonic) + $\met$ channel}

Here we present a search strategy for MSSM heavy Higgs bosons
in events with a leptonically decaying $\rm W$ boson 
($W \to \ell \nu, \ell = e, \mu$) and significant missing 
transverse energy for $\rm \sqrt{s} = 14\, TeV$ run of LHC with 
an integrated luminosity of 3000~$\ifb$. The $\rm W$ boson is 
produced through cascade decay of heavy Higgses, which first 
decays to a pair of neutralinos and/or charginos, which undergo 
further decay generating final states containing a $\rm W$ boson and 
large missing transverse energy. These events generally contain relatively 
soft leptons or jets. Here we follow the collider strategy as reported 
in Ref.~\cite{ATLAS:2014pna}, however some changes have been implemented in 
order to optimize the signal significances. The representative benchmark 
point BP-3 has been used to perform the detailed analysis. The 
decay chains of BP-3 relevant for this analysis are given below:
\begin{eqnarray}
p \, p \rightarrow H/A \rightarrow \chonepm\, \chtwomp, \quad \chtwomp \rightarrow W^{\mp}\,\lspone \nonumber \\
p \, p \rightarrow H/A \rightarrow \lspone\, \wt\chi_{2,3}^0, \quad  \wt\chi_{2,3}^0 \rightarrow W^{\pm}\,\chonemp. \nonumber  
\end{eqnarray}

From Table \ref{tab:mono_bpt}, it can be seen that for BP-3, $\lspone$ and 
$\chonepm$ are almost degenerate in mass, as a result, $\chonepm$ undergoes 
three-body decays resulting into soft leptons or jets in the final state. 
Similarly, the cascade decay chain originating from ${\rm H^{\pm}}$ can 
also lead to a mono-$W$ + $\met$ signature. However, we have not 
considered these decay chains in our analysis because of the 
relatively smaller production cross-sections of $H^{\pm}$.

The candidate events are required to have exactly one isolated electron 
or muon in the final state with $\rm p_{T} > 30~{\rm GeV}$. The 
candidate electron(muon) is required to satisfy $\rm |\eta| < 2.47(2.50)$. 
The isolation criteria for the candidate leptons is exactly same to what 
we had discussed in the last section. In addition, events are rejected 
if they contain one or more jets with $\rm p_{T} > 25$ GeV. For 
final state topologies with one isolated lepton and missing energy, 
we usually define a kinematic observable called transverse mass $M_T$ 
defined using the four momenta of the lepton and $\met$ as, 
\begin{eqnarray}
M_{T} = \sqrt{2 ~|p_{T}^{\ell}|~ \met (1 - \cos(\Delta \Phi_{\ell,\PMET}))} \nonumber
\end{eqnarray}
where, $\Delta \Phi_{\ell,\PMET}$ is the azimuthal angle between the 
candidate lepton and $\PMET$ direction. The dominant source of SM background is the 
$ p p \rightarrow \ell \nu j$ production channel which has been generated with 
a transverse mass cut $\rm M_{T}^{\ell \nu} > 100~{\rm GeV} $ where 
$\rm M_{T}^{\ell \nu} $ is the transverse mass of the lepton neutrino pair. 
Besides, processes like diboson production, $ t \bar{t}$ etc., also contribute 
significantly to the background list. Among the diboson modes, the dominant 
contribution comes from ${\rm W Z \rightarrow\, \ell\nu\nu\bar{\nu}}$ and 
${\rm W W \rightarrow \ell \nu \ell \nu }$ ($\ell = e,\mu$) channels. 

\begin{table}[!htb]
\begin{center}
\begin{tabular}{|c|c|}
\hline
 Signal Regions & Selection Cuts \\ 
\hline
SRA2 & ${\met > 50~{\rm GeV}} \quad \& \quad M_{T} > 175~{\rm GeV} $ \\
SRB2 & ${\met > 100~{\rm GeV}} \quad \& \quad M_{T} > 125~{\rm GeV} $ \\
\hline
\end{tabular}
\caption{Signal regions for the mono-$W + \met$ analysis. The transverse 
mass ($M_T$) is defined using the four momenta of the lepton and the missing 
transverse momentum $\met$.}
\label{tab:SRs2}
\end{center}
\end{table}

Similar to the earlier analysis, two signal regions, SRA2 and SRB2, with 
different optimized values of $\met$ and $M_{T}$, are constructed, as 
displayed in Table \ref{tab:SRs2}. For both the signal regions SRA2 and SRB2, we further 
selects those leptons which satisfy $|\eta^{\ell}| < 1.5$ with $\ell = e,\mu$. 
The choice of these signal regions is driven from the optimization of the 
signal significances. We investigate several combinations of selection variables 
$\met$ and $M_{T}$, and among them, two signal regions (SRA2 and SRB2) are 
chosen which yield the most efficient optimization of signal 
significances. The expected number of signal and background events 
at the 14 TeV run of LHC with 3000~$\ifb$ of luminosity are shown in 
Table \ref{tab:monow_bg} for both the above-mentioned signal regions. Similar to our 
previous analysis, one needs to restrict the lower bounds on $\met$ to 150 GeV 
as the signal yields become relatively insignificant as one shifts towards 
high $\met$ regime. Here again, we calculate the statistical significance using 
the signal and background events, and find that it is around $2\sigma$ with 3000~${\ifb}$ luminosity 
at the 14 TeV run of LHC. This indicates that it would be very difficult to discover/exclude the 
heavy Higgs through this channel. More efficient signal optimizations and more 
precise understanding of the backgrounds may be required to make this search 
channel more efficient.

\begin{table}[!h]
\begin{center}
\begin{tabular}{|c|c|c c c c c|c|}
\hline
Signal & Signal & \multicolumn{5}{c|}{Backgrounds} & Significance \\
\cline{2-8}
Regions   & BP-3   & $ l \nu $  & $ W W $ & $ W Z $ & $ t \bar{t}$ & $ Z Z $ & BP-3 \\
\cline{1-8}
      SRA2   & 6572  & $1.0\cdot 10^{7}$ &  76427   &  68426  & 58204  & 9088 & 2.05 \\
      SRB2   & 5499  & $6.8\cdot 10^{6}$ &  47603  &  52266   & 59380  & 8071 & 2.07 \\
\hline
\end{tabular}
\caption{\small \it Number of signal and background events at the 14 TeV run 
of LHC with 3000~$\ifb$ of luminosity obtained after the imposition of our selection cuts 
for the mono-$W+\met$ channel. Here we focus on the third representative 
benchmark point BP-3. }
\label{tab:monow_bg}
 \end{center}
 \end{table}



\subsection{Trilepton + $\met$ channel }
\label{sec:3l}

In this section, we present a search strategy for the heavy Higgses 
$H,A,H^{\pm}$ in events with leptonically decaying W and Z bosons 
($W \to \ell \nu, Z \to \ell\ell, \ell= e,\mu$) 
produced through cascade decay of heavy Higgses with significant missing 
transverse energy at $\sqrt{s} = 14~{\rm TeV}$ run of LHC with an 
integrated luminosity of 3000~$\ifb$. Here we follow the collider 
strategy reported in Ref.~\cite{CMStrilep}, and consider the 
representative benchmark point BP-4. The relevant decay chains 
which generate the above mentioned signature are given below:
\begin{eqnarray}
p\,p \rightarrow H/A,\quad H/A \rightarrow \chonepm\, \chtwomp,\quad \chtwomp \rightarrow W^{\mp}\, \lspthree,  \quad \lspthree \rightarrow Z \, \lspone  \nonumber \\
p\,p \rightarrow H/A,\quad H/A \rightarrow \lspthree\, \lspfour,\quad \lspthree \rightarrow Z \, \lspone,\quad \lspfour \rightarrow W^{\pm}\,\chonemp, \nonumber
\end{eqnarray}

In BP-4, $\chonepm$ undergoes only three body decay resulting in 
additional leptons or jets in the final state. The cascade decay chain 
originating from a charged Higgs, with subsequently decays to a 
chargino-neutralino pair, followed by the decay of charginos and neutralinos to 
W boson + $\met$ and Z boson + $\met$, respectively, gains a significant 
branching fraction and leads to a trilepton plus $\met$ signature. 
However, because of low production cross-section of $H^{\pm}$, 
the resultant contribution of these decay modes to the trilepton signatures 
is relatively low. As a result, similar to previous analyses, we have not 
taken into account the charge Higgs cascade decay chains into our analysis.

The event selection criteria are almost similar to our previous analyses. 
Candidate events are required to have exactly three isolated leptons (electron/muon) 
with $\rm p_{T} > 20$ GeV. The electrons (muons) are required to 
lie within $|\eta| < 2.47 (2.5)$. Isolation criteria discussed in the 
previous section applies here as well. Among the three candidate leptons, 
there must be at least one SFOS pair with invariant mass in the range 
$\rm |M_{Z} \pm 15|$~GeV. If there are more than one SFOS pairs satisfying 
the previous condition, then the lepton pair with invariant mass closest 
to the Z boson mass is finally identified as the SFOS pair. The transverse 
mass $M_{T}$ is defined with respect to the lepton which is not a 
part of the SFOS pair.

\begin{table}[!htb]
\begin{center}
\begin{tabular}{|c|c|}
\hline
 Signal Regions & Selection Cuts \\ 
\hline
SRA2 & ${\met > 50~{\rm GeV}} \quad \& \quad M_{T} > 150~{\rm GeV}$ \\
SRB2 & ${\met > 50~{\rm GeV}} \quad \& \quad M_{T} > 200~{\rm GeV} $ \\
\hline
\end{tabular}
\caption{Signal regions for the $3l + \met$ analysis. The transverse 
mass ($M_T$) is defined using the four momenta of the lepton not forming the SFOS pair and the missing 
transverse momentum $\met$.}
\label{tab:SRs3}
\end{center}
\end{table}

The SM backgrounds for this search channel are $\rm WZ \rightarrow \ell \nu \ell \ell$ 
and $\rm ZZ \rightarrow 4\ell$ processes. Here also, we define two signal regions, 
SRA3 and SRB3, for different conditions of $\met$ and $M_{T}$, as displayed in 
Table \ref{tab:SRs3}. The choice of signal regions are motivated from 
optimization of signal efficiencies as well as the experimental analysis. The number of 
signal and background events expected at the 14 TeV run of LHC with 3000~$\ifb$ of luminosity 
are listed in Table \ref{tab:3l_bg}. The estimated statistical 
significances are very small in this search channel, which indicates that 
probing the MSSM heavy Higgses through trilepton + $\met$ channel, 
with leptons originating from the cascade decays of the heavy Higgses, 
would be an extremely challenging task even at 14 TeV run of LHC.   

\begin{table}[!htb]
\begin{center}
\begin{tabular}{|c|c|c c|c|}
\hline
Signal & Signal & \multicolumn{2}{c|}{Backgrounds} & Significance \\
\cline{2-5}
Regions   & BP-4   & $ W Z $ & $ Z Z $ & BP-4 \\
\cline{1-5}
      SRA3   & 6.92  & 544 & 21   &  0.29   \\
      SRB3   & 4.33  & 389 & 16  &  0.22    \\
\hline
\end{tabular}
\caption{\small \it Number of signal and background events at the 14 TeV run 
of LHC with 3000~$\ifb$ of luminosity obtained after the imposition of our selection cuts 
for the $3l+\met$ channel. Here we focus on the fourth representative 
benchmark point BP-4. }
\label{tab:3l_bg}
 \end{center}
 \end{table}




\section{Summary}
\label{summary}

Precise measurements of the properties of the discovered Higgs boson has been 
one of the major goals of the LHC physics program. So far, LHC 7 and 8 TeV data 
reveals that the properties of this new particle are consistent with 
the SM Higgs boson within the uncertainty in Higgs couplings 
measurements. Many new physics models beyond the SM contain additional 
Higgs doublets leading to additional Higgses. A multitude of searches 
have been performed by the ATLAS and CMS collaborations to probe the heavy 
Higgses (H, A and $\rm H^{\pm}$) of the MSSM 
through their decay to the SM particles. Note that so far none of these 
searches have provided a clear signature of the heavier Higgs states. In this 
regard, it becomes important to examine the non-SM decay modes of the 
heavier Higgses. The primary motivation for performing such analyses is that 
when these heavy Higgses decay to light SUSY particles, all the branching 
ratios to SM particles acquire significant modifications, thereby changes the whole framework 
of the LHC search strategy. Moreover, there exists certain regions 
of the parameter space with intermediate $\tan\beta$ ($\sim$ 5-15), 
where the heavy Higgs couplings to the SM particles become small. 
However, in this region of interest, one can obtain appreciable 
amount of non-SM decays which can be studied at the LHC. One such non-SM 
decay mode could be the decay of the heavy Higgses to the light 
electroweakinos (charginos and neutralinos). Study of these non-SM 
decay modes in the light of updated LHC data is precisely 
the goal of this paper.

In the presence of light SUSY particles, and if kinematically allowed, 
these heavy Higgs bosons can decay to SUSY particles with a 
significantly high branching fraction. In order to estimate these 
non-SM decays, we start with a simple-minded scan making 
the sparticle sector sufficiently heavy except the electroweakinos, 
which we assume sufficiently light such that Higgs decay is kinematically 
allowed, and vary the $\tan\beta$ over a wide range. From this simple 
analysis, we find that these non-SM `ino' decays can be as large as 
70-80\% for relatively low values ($\sim$ 5--10) of $\tan\beta$. 
However, here we do not impose the updated LHC Higgs data and also 
other low energy flavour data. So, in order to perform a detailed 
analysis, we scan the MSSM parameter space with the parameters 
relevant to the electroweakinos, namely $M_{1},M_{2},\mu~{\rm and} \tan\beta$, 
keeping the sleptons and squarks fixed at high scale. All the scanned 
points are required to have the lightest Higgs boson mass in the range 
122 GeV to 128 GeV. We also consider the constraints from the Higgs 
couplings measurements, where we use the 95\% C.L. contours obtained 
from a global fit analysis performed by the ATLAS and CMS collaborations. 
In addition, updated bounds from low energy flavour data in terms on rare b-decays 
$\rm BR(b \rightarrow s\gamma)$ and $\rm BR(B_{s} \rightarrow \mu^{+}\mu^{-})$ 
are also considered in our analysis. Now, depending on the values and 
hierarchy of the electroweakino parameters, we construct ten 
representative models Model-B to Model-BWH, and then estimate various 
non-SM `ino' decay modes of the heavy Higgs bosons. From the scan, we 
observe that some of these ino-modes can be as large as 35--40\% 
even after satisfying the updated LHC data. These non-SM decay modes 
crucially depends on the gaugino-higgsino mixing, or precisely 
on the composition of these electroweakino states. However, as we have 
already mentioned, both the ATLAS and CMS collaborations have 
searched for additional Higgses at the LHC and put bounds on masses and 
couplings of these heavy Higgses. So our next task should be to check 
whether these heavy Higgses, whose non-SM decay modes we are calculating, 
are still allowed by the current data. Moreover, we also require to perform 
consistency checks of these light electroweakinos with the LHC direct search 
bounds on their masses and couplings obtained at the end of run-I. 
Instead on implementing the current bounds from the direct searches of 
the heavy Higgses and electroweakinos on the each points corresponding 
to our scanned data set, we choose four representative benchmark 
points making sure that all these points satisfy the current 
LHC run-I and run-II data.        

Once we select those benchmark points consistent with the present LHC data, 
we ask this simple question, can we utilize this large branching ratios of 
the heavy Higgses to electroweakinos and look for collider 
signatures of the same through cascade decay at the 14 TeV high luminosity 
run of the LHC ? So, we perform dedicated collider analysis at the 
$\rm \sqrt{s} =14~{\rm TeV}$ corresponding to an integrated luminosity of 
$\rm 3000~\ifb$. We focus on the leptonic modes of the cascade decays 
as these channels are very clean at the LHC busy environment and also we have 
better control over the backgrounds. Our collider analysis can be divided 
into two parts, one focusing on the mono-X plus missing energy signatures 
where X represents W or Z bosons, and another using the trileptonic channel with 
significant amount of missing energy. Among the four representative 
benchmark points, BP-1 and BP-2 are focused to probe the $H,A$ through 
the mono-Z + $\met$ signature, while BP-3 and BP-4 are used to study 
the heavy Higgses through the mono-W + $\met$ and trilepton + $\met$ signatures 
respectively. Following an ATLAS study and making suitable changes in the 
selection cuts we find that the mono-Z channel has the best sensitivity to 
probe these heavy Higgses while other two modes posses mild 
sensitivity for the exclusion of these additional Higgses at the 14 TeV 
run of LHC with 3000 $\ifb$ luminosity.  

Before we end, we would like to note few important issues. In this work, 
we focus on the decay of the heavy MSSM Higgses to the light electroweakinos 
only setting other SUSY particles decoupled from the spectrum. However, in 
principle, some of these sparticles, say top squarks, tau sleptons etc., can 
be also light, thereby those decay modes will also contribute to the non-SM 
decay of the heavy Higgses. Moreover, in our collider analysis we restrict 
ourselves within the leptonic modes only, however plethora of final states 
involving leptons and jets are possible. Furthermore, some of the final state 
particles, say $W,Z$ bosons or Higgs bosons, can acquire large transverse 
momentums, in those situation one need to invoke the state-of-the-art jet 
substructure techniques to improve the sensitivity of these heavy resonance 
searches. All of these points are beyond the scope of the current paper, we leave 
these issues for a future correspondence focussing on the high luminosity runs of the 
LHC.

\small
\baselineskip 16pt
{\large \bf \underline {Acknowledgements:}}
\vskip 0.12cm
Work of B. Bhattacherjee is supported by Department of Science and Technology, Government
of INDIA under the Grant Agreement numbers IFA13-PH-75 (INSPIRE Faculty Award). 
The work of A. Choudhury  is supported by the Lancaster-Manchester-Sheffield 
Consortium for Fundamental Physics under STFC Grant No. ST/L000520/1. 



\end{document}